\documentclass[10pt]{article} 

\usepackage[preprint]{rlj} 

\usepackage{amssymb}            
\usepackage{mathtools}          
\usepackage{mathrsfs}           
\usepackage{enumitem}
\usepackage{graphicx}           
\usepackage{subcaption}         
\usepackage[space]{grffile}     
\usepackage{url}                
\usepackage{lipsum}             

\usepackage{pgfplots}
\pgfplotsset{compat=1.18}

\usepackage{booktabs}
\usepackage{multirow}
\usepackage{makecell}
\usepackage{colortbl}

\definecolor{tablegray}{RGB}{248,248,248}
\definecolor{bestblue}{RGB}{235,245,255} 

\newcommand{\best}[1]{\cellcolor{bestblue}\textbf{#1}}

\title{SCoUT: Scalable Communication via Utility-Guided Temporal Grouping in Multi-Agent Reinforcement Learning}

\setrunningtitle{SCoUT}

\author{Manav Vora, Gokul Puthumanaillam, Hiroyasu Tsukamoto, Melkior Ornik}

\emails{{mkvora2, gokulp2, hiroyasu, mornik}@illinois.edu}

\affiliations{
\textbf{University of Illinois Urbana-Champaign}
}

\contribution{
\textbf{SCoUT}, a scalable learned-communication method that resamples {soft} agent groups every $K$ environment steps and uses the induced affinity as a differentiable {bias} over per-step recipient selection. This \emph{temporally-extended} soft-grouping mechanism turns per-step combinatorial link selection over $2^{N(N-1)}$ possible directed communication graphs into structured, differentiable routing guided by group affinities during training.
}{
Prior selective communication methods typically re-decide communication structure at every step via scheduling \citep{kim2019schednet}, attention-based routing \citep{Das2019TarMAC}, or imposing a global topology \citep{Li2025ExpoComm}; SCoUT instead introduces a slowly varying latent structure that guides routing without hard constraints.
}

\contribution{
A \emph{group-aware critic} that predicts group-level values from global information and maps them to per-agent value baselines through the soft grouping assignments, reducing critic output complexity and stabilizing CTDE training in large populations.
}{
This differs from monolithic centralized critics in standard CTDE baselines \citep{yu2021mappo} and is complementary to value-factorization methods \citep{rashid2020monotonic} because the factorization is tied to the learned communication structure in our setting.
}

\contribution{
A \emph{counterfactual mailbox} credit-assignment rule that yields learning signals for both the binary send action and recipient selection by comparing the recipient’s predicted value under the actual mailbox versus a leave-one-out (counterfactual) mailbox.
}{
Inspired by counterfactual baselines in multi-agent actor-critic learning \citep{foerster2018coma}, but specialized to communication by isolating the marginal contribution of individual messages.
}

\contribution{
Empirical evaluations on large-population benchmarks showing that SCoUT can be trained efficiently in scenarios with hundreds of agents and improves performance over learned-communication baselines as population size grows. 
}
{
We consider scenarios with roughly 1.6$\times$ and 12$\times$ more controlled agents than prior work in Battle and Pursuit respectively. 
}


\summary{
Learning communication can improve coordination in partially observed MARL, but scaling to large teams is difficult because (i) deciding \emph{when} and \emph{with whom} to communicate induces many coupled discrete choices and (ii) the effect of any single message is hard to isolate for learning. We propose \textbf{SCoUT} (\textbf{S}calable \textbf{Co}mmunication via \textbf{U}tility-guided \textbf{T}emporal grouping), a scalable learned-communication method that introduces temporal and agent abstraction for learned communication. SCoUT resamples soft agent groups every $K$ steps to induce a differentiable affinity that biases recipient selection, while a group-aware critic predicts group-level values and maps them to per-agent baselines to reduce critic complexity. To obtain precise learning signals for communication, SCoUT uses counterfactual mailbox-based advantages that isolate the contribution of individual messages for both send and recipient decisions. Experiments on benchmarks show that SCoUT remains effective as the agent population grows to hundreds, whereas prior learned-communication baselines degrade with scale.
}

\begin{document}

\makeCover  
\maketitle  

\begin{abstract}
Communication can improve coordination in partially observed multi-agent reinforcement learning (MARL), but learning \emph{when} and \emph{who} to communicate with requires choosing among many possible sender-recipient pairs, and the effect of any single message on future reward is hard to isolate. We introduce \textbf{SCoUT} (\textbf{S}calable \textbf{Co}mmunication via \textbf{U}tility-guided \textbf{T}emporal grouping), which addresses both these challenges via temporal and agent abstraction within traditional MARL. During training, SCoUT resamples \textit{soft} agent groups every \(K\) environment steps (macro-steps) via Gumbel-Softmax; these groups are latent clusters that induce an affinity used as a differentiable prior over recipients. Using the same assignments, a group-aware critic predicts values for each agent group and maps them to per-agent baselines through the same soft assignments, reducing critic complexity and variance. Each agent is trained with a three-headed policy: environment action, send decision, and recipient selection. To obtain precise communication learning signals, we derive counterfactual communication advantages by analytically removing each sender's contribution from the recipient's aggregated messages. This counterfactual computation enables precise credit assignment for both send and recipient-selection decisions. At execution time, all centralized training components are discarded and only the per-agent policy is run, preserving decentralized execution. Experiments on large-scale benchmarks show that SCoUT learns targeted communication and remains effective in large scenarios, while prior methods degrade as the population grows. Finally, ablations confirm that temporal grouping and counterfactual communication credit are both critical for scalability. 

Project website, videos and code: \hyperlink{https://scout-comm.github.io/}{https://scout-comm.github.io/}
\end{abstract}

\section{Introduction}
Communication is a natural mechanism for coordination in partially observed multi-agent reinforcement learning (MARL): exchanging local information or intent can mitigate limited individual views and help synchronize behavior toward a shared objective \citep{bernstein2002complexity,hernandezleal2019survey,busoniu2008survey}. Accordingly, recent work learns communication protocols and routing decisions end-to-end alongside agent policies, rather than prescribing a fixed topology or hand-designed messages \citep{niu2021multi,li2021dicg}. Yet scaling learned communication beyond small teams remains difficult, with growing computational cost and training instability as the agent population increases \citep{li2021dicg}.

A key bottleneck in scaling learned communication is that deciding \emph{when} and \emph{with whom} to communicate is inherently combinatorial: at each step, the team implicitly chooses a communication graph over $N$ agents, yielding $2^{N(N-1)}$ possible link patterns. Fully connected architectures avoid explicit routing decisions but require $O(N^2)$ message exchanges and aggregations per step, which can become costly and inject substantial irrelevant traffic as $N$ grows \citep{sukhbaatar2016commnet,foerster2016dial}. Selective communication methods reduce per-step communication and computation by gating or scheduling transmissions \citep{Singh2019IC3Net,kim2019schednet} or by learning targeted routing with attention \citep{jiang2018atoc,Das2019TarMAC,Liu2020G2ANet}. 
However, these approaches are typically trained under centralized training with decentralized execution using actor-critic objectives, where centralized value functions provide learning signals for both environment actions and communication decisions \citep{lowe2017maddpg,yu2021mappo}. As the population grows, two challenges often persist at training time: (i) \emph{training-time scalability}--critics and communication modules must process high-dimensional, multi-agent information; and (ii) \emph{credit assignment}--attributing downstream returns to specific messages and recipient choices when many communications jointly influence outcomes \citep{foerster2018coma}. 
Empirically, learned-communication results are typically demonstrated at scales of $\mathcal{O}(10$-$100)$ agents \citep{Li2025ExpoComm}.

To make learned communication feasible for \emph{large teams} with hundreds of agents, we propose \textbf{SCoUT} (\textbf{S}calable \textbf{CO}mmunication via \textbf{U}tility-guided \textbf{T}emporal grouping), a training framework that introduces \emph{temporal} and \emph{agent} abstraction. The core idea is to treat the communication structure as a slowly varying latent variable: every $K$ environment steps (a macro-step), SCoUT resamples a set of soft agent groups--a differentiable clustering of agents to a small number of latent groups--using differentiable categorical sampling \citep{jang2017gumbel,maddison2017concrete}. These group affinities induce a differentiable prior over recipient selection and message aggregation in a one-step-latency \emph{mailbox}, i.e., a per-agent message buffer of received messages from the previous step. The same group assignments also factor value estimation through a group-aware critic that predicts group-level values and maps them back to per-agent baselines, reducing critic complexity and variance. We train each agent with a three-headed PPO policy (environment action, send, recipient selection) \citep{schulman2017ppo,schulman2016gae}, and derive counterfactual communication advantages by analytically removing a sender's contribution from a recipient's aggregated mailbox, in the spirit of counterfactual baselines for multi-agent credit assignment \citep{foerster2018coma}.

\paragraph{Contributions.}
\begin{enumerate}[label=(\roman*), leftmargin=*, itemsep=0pt, topsep=0pt, parsep=0pt, partopsep=0pt]
    \item We introduce a \emph{temporally-extended} soft-grouping mechanism for MARL communication, that induces a learned distribution over recipients. This mechanism replaces per-step combinatorial link selection over $2^{N(N-1)}$ possible directed communication graphs, with structured, differentiable routing guided by group affinities.
    \item We propose a \emph{group-aware critic} that factors value prediction through soft groups and maps group values back to per-agent baselines, improving the scalability and stability of CTDE training in large populations.
    \item We develop a \emph{counterfactual mailbox} credit-assignment rule that yields direct learning signals for both sending and recipient selection by analytically removing a sender's contribution from a recipient's aggregated messages.
\end{enumerate}

At execution time, SCoUT discards centralized components (group sampler, critic, and counterfactual computations) and runs only the learned per-agent policy, preserving decentralized execution. 

We evaluate SCoUT on large-scale cooperative benchmarks from PettingZoo \citep{terry2021pettingzoo} and MAgent \citep{zheng2018magent}, where it learns targeted communication while scaling to hundreds of agents and outperforming state-of-the-art baselines as population size grows; ablations in Pursuit isolate the impact of temporal grouping and counterfactual credit assignment.

\section{Related Work}

\paragraph{Foundations and CTDE training.}
Partially observed multi-agent decision-making is captured by decentralized control formalisms such as Dec-POMDPs \citep{bernstein2002complexity} and by Markov games \citep{littman1994markov}. Deep MARL commonly adopts centralized training with decentralized execution (CTDE), using global information to train critics or value decompositions while executing decentralized policies \citep{lowe2017maddpg,foerster2018coma}. Representative CTDE methods include centralized critics (MADDPG) \citep{lowe2017maddpg}, counterfactual actor--critic learning (COMA) \citep{foerster2018coma}, and value factorization (VDN/QMIX, QTRAN, QPLEX) \citep{sunehag2018vdn,rashid2020monotonic,son2019qtran,wang2021qplex}; PPO-style CTDE baselines such as MAPPO are also competitive \citep{schulman2017ppo,yu2021mappo}. As the number of agents grows, however, centralized critics can become a bottleneck: they must summarize high-dimensional interactions and provide low-variance learning signals. SCoUT addresses this by factoring value estimation through soft groups and mapping group values to per-agent baselines.

\paragraph{Learned communication in MARL.}
A long line of work learns differentiable communication end-to-end, enabling agents to exchange messages that improve coordination under partial observability. CommNet aggregates broadcast-style continuous messages \citep{sukhbaatar2016commnet}, while DIAL/RIAL learn communication alongside decentralized policies with differentiable and discrete channels \citep{foerster2016dial}. For large populations, dense communication becomes inefficient and noisy, and message utility is difficult to disentangle. Selective schemes reduce transmissions by learning \emph{when} to communicate (IC3Net) \citep{Singh2019IC3Net} or scheduling a subset of speakers under bandwidth constraints \citep{kim2019schednet}. Yet these methods typically re-decide communication each step and rely on centralized training signals, making credit assignment for communication increasingly challenging at scale. To alleviate this scalability issue, SCoUT introduces temporal abstraction via soft groups that persist for $K$ steps and a counterfactual mailbox signal that efficiently trains both sending and recipient selection.

\begin{figure*}[h]
    \centering
    \includegraphics[width=0.85\linewidth]{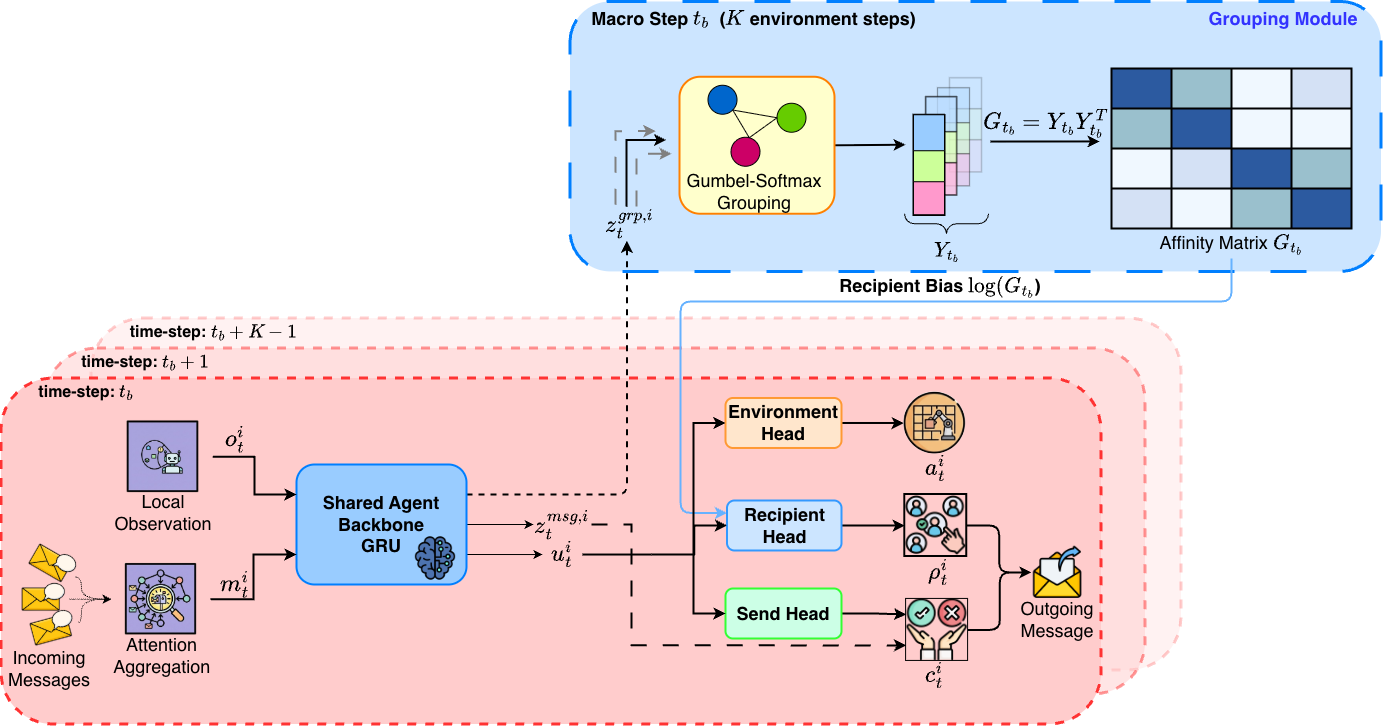}
    \caption{SCoUT forward-pass overview across two timescales. At each macro-step boundary $t_b$, the grouping module samples soft assignments $Y_{t_b}$ and forms an affinity matrix $G_{t_b}=Y_{t_b}Y_{t_b}^\top$, which is held fixed over the subsequent $K$ primitive steps and used as a log-bias $\log(G_{t_b})$ for recipient selection. At each primitive step $t$, each agent embeds its local observation $o_t^i$ and mailbox input $m_t^i$, updates a shared GRU backbone, and outputs a three-headed policy: environment action $a_t^i$, send decision $c_t^i$, and recipient $\rho_t^i$. If $c_t^i=1$, the agent transmits message content $x_t^i=z_t^{\mathrm{msg},i}$ to the chosen recipient; recipients aggregate incoming messages into the next-step mailbox.}
    \label{fig:scout_overview}
\end{figure*}

\paragraph{Targeted routing, attention, and structure.}
Beyond ``speak or stay silent,'' targeted routing methods learn \emph{with whom} to communicate. ATOC forms temporary coordination groups via attention \citep{jiang2018atoc}, TarMAC learns attention-based addressing \citep{Das2019TarMAC}, and G2ANet models relations with multi-stage attention \citep{Liu2020G2ANet}; ToM2C incorporates Theory-of-Mind-style inference \citep{wang2021tom2c}. These approaches provide richer structure than broadcasting but often evaluate routing over many potential recipients and can be expensive or unstable when recomputed at every step for large $N$. In parallel, topology-design approaches advocate fixed global structures; for example, ExpoComm leverages exponential-graph topologies for small-diameter dissemination \citep{Li2025ExpoComm}. SCoUT sits between these directions by learning a task-adaptive grouping structure that changes slowly over time and acts as a structured bias over recipient selection, without fixed topology or per-step global rerouting.

\paragraph{Scalability and many-agent benchmarks.}
Scaling MARL has also been pursued via mean-field approximations that replace many-body interactions with aggregate effects \citep{yang2018meanfield}. Standardized benchmarks enable evaluation across regimes, including SMAC \citep{samvelyan2019smac} and many-agent gridworlds in MAgent \citep{zheng2018magent}. SCoUT is motivated by large \emph{finite} populations with heterogeneous local observations and structured coordination needs, where dense communication and monolithic critics become impractical and credit assignment for communication decisions is particularly acute.

\section{Problem Setting}
\label{sec:problem_setting}
We consider cooperative multi-agent reinforcement learning under partial observability, modeled as a decentralized partially observable Markov decision process (Dec-POMDP) \citep{bernstein2002complexity}. A Dec-POMDP is defined by
$\langle \mathcal{S}, \{\mathcal{A}_i\}_{i=1}^N, \mathcal{P}, r, \{\mathcal{O}_i\}_{i=1}^N, \mathcal{Z}, \gamma \rangle$,
where $s_t\in\mathcal{S}$ is the environment state, each agent $i\in\{1,\dots,N\}$ selects an environment action $a_t^i\in\mathcal{A}_i$, the transition kernel is $\mathcal{P}(s_{t+1}\mid s_t,a_t^1,\dots,a_t^N)$, and all agents share a cooperative reward $r(s_t,a_t^1,\dots,a_t^N)$ with discount factor $\gamma\in(0,1)$. Each agent receives a local observation $o_t^i\in\mathcal{O}_i$ emitted by $\mathcal{Z}(o_t^1,\dots,o_t^N\mid s_t)$ and acts based on its local history $h_t^i=(o_{1:t}^i,a_{1:t-1}^i)$.

\paragraph{Communication interface.}
In addition to environment actions, agents may communicate over a directed, point-to-point channel. At time $t$, agent $i$ chooses (i) a binary send decision $c_t^i\in\{0,1\}$, (ii) if $c_t^i=1$, a recipient index $\rho_t^i\in\{1,\dots,N\}$, and (iii) a message vector $x_t^i\in\mathcal{X}\subseteq\mathbb{R}^d$ with fixed dimension $d$. We assume a synchronous \emph{one-step latency}: messages sent at time $t$ are available to recipients at time $t{+}1$. Since a recipient may receive multiple messages in the same step, we represent its incoming communication by a fixed-dimensional \emph{mailbox} vector $m_{t+1}^j\in\mathbb{R}^{d_m}$ obtained via an aggregation operator
\begin{equation*}
m_{t+1}^j \;=\; g\!\left(\{x_t^i:\; c_t^i=1,\ \rho_t^i=j\}\right),
\end{equation*}
where $g$ maps a multiset of incoming messages to a single vector and $m_{t+1}^j=\mathbf{0}$ if no messages are addressed to $j$.

\paragraph{Policies and objective.}
A stochastic decentralized policy for agent $i$ is a conditional distribution over its decisions given its local information, i.e.,
$\pi^i(\cdot\mid h_t^i,m_t^i)$.
We assume parameter sharing and use a single policy $\pi_\theta$ for all agents \citep{christianos2021scaling,terry2020revisiting}:
\begin{equation*}
(a_t^i, c_t^i, x_t^i, \rho_t^i) \sim \pi_{\theta}(\cdot \mid h_t^i, m_t^i), \qquad \forall i \in \{1,\dots,N\}.
\end{equation*}
This induces a joint policy that factorizes across agents conditioned on their local inputs,
\begin{equation*}
\boldsymbol{\pi}_{\theta}(\mathbf{a}_t, \mathbf{c}_t, \mathbf{x}_t, \boldsymbol{\rho}_t \mid \mathbf{h}_t, \mathbf{m}_t)
= \prod_{i=1}^N \pi_{\theta}(a_t^i, c_t^i, x_t^i, \rho_t^i \mid h_t^i, m_t^i).
\end{equation*}
Our goal is to maximize the expected discounted team return
\begin{equation}
J(\boldsymbol{\pi}_{\theta})=
\mathbb{E}_{\boldsymbol{\pi}_{\theta}}\!\left[\sum_{t=0}^{\infty}\gamma^t\, r(s_t, a_t^1,\dots,a_t^N)\right].
\end{equation}

\paragraph{Learning setup (CTDE).}
We adopt centralized training with decentralized execution (CTDE): at execution time, each agent acts using only $(h_t^i,m_t^i)$, while during training we may leverage additional global information (which in our case is the global state) to learn critics or auxiliary components \citep{lowe2017maddpg,foerster2018coma}.

\section{SCoUT}
\label{sec:scout}
We now describe our framework for scalable, utility-guided communication in cooperative MARL. SCoUT (i) introduces a slowly varying latent communication structure via soft grouping rather than re-deciding a dense communication graph at every step, (ii) uses this structure to simplify and factor value estimation through a group-aware critic, and (iii) trains targeted communication decisions, including the binary send action $c_t^i\in\{0,1\}$ and recipient choice $\rho_t^i$, using counterfactual advantages that attribute utility to individual messages. Figure~\ref{fig:scout_overview} provides a training forward-pass overview of SCoUT, highlighting the macro-step grouping and the per-step three-headed communication policy.

\subsection{Agent backbone and descriptors}
\label{sec:scout_repr}

Each agent runs a shared recurrent backbone to maintain memory under partial observability.
Let $\varphi(\cdot)$ embed the concatenated local observation and mailbox input.
We maintain an internal recurrent state $u_t^i$ with a GRU \citep{cho2014learning}:
\begin{equation*}
u_t^i = \mathrm{GRU}\!\bigl(\varphi([o_t^i,\, m_t^i]),\, u_{t-1}^i\bigr).
\end{equation*}
From $u_t^i$ we compute a lightweight descriptor via two learned projections:
(i) a \emph{grouping} descriptor $z_{t}^{\mathrm{grp},i}$, evaluated only every $K$ steps (at macro-step boundaries), and
(ii) a \emph{message} descriptor $z_{t}^{\mathrm{msg},i}$, used as message content when the agent transmits.
All representation parameters are trained end-to-end through gradients from the policy and value objectives, as well as the grouping and communication components introduced below.

\subsection{Temporal soft grouping and affinity prior}
\label{sec:scout_grouping}
A central difficulty in large-team communication is that choosing a directed communication pattern at every step introduces many coupled discrete decisions; with sparse/delayed rewards, this can lead to high-variance gradients and brittle credit assignment. SCoUT mitigates this by introducing a slowly varying latent structure: every macro-step $b$ (every $K$ primitive steps), we sample {soft groups} that act as (i) a coarse, persistent notion of communication neighborhoods and (ii) a reusable structure for downstream components (see the top panel of Figure~\ref{fig:scout_overview}).

Let $t_b$ denote the first primitive step of macro-step $b$. We assign each agent $i$ to one of $M\ll N$ latent groups using its grouping descriptor $z_{t_b}^{\mathrm{grp},i}$ and a set of learnable group prototypes $\{p_g\}_{g=1}^M$. Intuitively, prototypes serve as anchor points, akin to soft $k$-means centroids. We compute assignment logits via cosine similarity:
\begin{equation*}
\ell_{t_b}^{i,g}
= \alpha \cdot
\frac{\langle z_{t_b}^{\mathrm{grp},i},\, p_g \rangle}
{\|z_{t_b}^{\mathrm{grp},i}\|\,\|p_g\|},
\qquad g \in \{1,\dots,M\},
\end{equation*}
where $\alpha>0$ is a learnable scale. We then sample \emph{soft} assignments with the Gumbel--Softmax reparameterization \citep{jang2017gumbel,maddison2017concrete}:
\begin{equation*}
Y_{t_b}^{i,g}
=
\mathrm{softmax}\!\left(\frac{\ell_{t_b}^{i,\cdot} + \epsilon^{i,\cdot}}{\tau}\right)_g,
\quad
\epsilon^{i,g} \sim \mathrm{Gumbel}(0,1),
\end{equation*}
with temperature $\tau$, yielding $Y_{t_b}\in\mathbb{R}^{N\times M}$. Following common practice in differentiable pooling  \citep{ying2018diffpool}, we form an affinity matrix
\begin{equation*}
G_{t_b} \;=\; Y_{t_b}Y_{t_b}^\top \in \mathbb{R}^{N\times N}.
\end{equation*}
where $G_{t_b}^{ij}$ is large when agents $i$ and $j$ are assigned to similar groups.This affinity is held fixed for the entire macro-step and serves as a differentiable prior over communication partners.

\paragraph{Grouping objective.}
Because grouping is a stochastic latent decision, we optimize it with a score-function objective \citep{williams1992reinforce} using a macro-step advantage signal. Let $P_\tau^{i,g}=\mathrm{softmax}(\ell_{t_b}^{i,\cdot}/\tau)_g$ denote the assignment probabilities and let $A_{b}^{\mathrm{grp},i}$ denote the macro-step advantage produced by the group-aware critic (Section~\ref{sec:scout_critic}). We minimize
\begin{equation*}
\mathcal{L}_{\mathrm{grp}}
=
\mathcal{L}_{\mathrm{grp}}^{\mathrm{PG}}
+\lambda_{\mathrm{bal}}\mathcal{L}_{\mathrm{grp}}^{\mathrm{bal}}
+\lambda_{\mathrm{ent}}\mathcal{L}_{\mathrm{grp}}^{\mathrm{ent}}
+\lambda_{\mathrm{edge}}\mathcal{L}_{\mathrm{grp}}^{\mathrm{edge}},
\end{equation*}
where $\lambda_{\mathrm{bal}},\lambda_{\mathrm{ent}},\lambda_{\mathrm{edge}}\ge 0$ are user-chosen coefficients controlling the strength of each regularizer. The policy-gradient term is
\begin{equation*}
\mathcal{L}_{\mathrm{grp}}^{\mathrm{PG}}
=
-\sum_{b}\sum_{i=1}^N
A_{b}^{\mathrm{grp},i}\,
\sum_{g=1}^M Y_{t_b}^{i,g}\,\log P_\tau^{i,g},
\end{equation*}
which applies the score-function weight to the sampled assignment.
We additionally regularize against degenerate or premature collapse:
\begin{align*}
\mathcal{L}_{\mathrm{grp}}^{\mathrm{bal}}
&=
\left\|
\frac{1}{N}\sum_{i=1}^N P_\tau^{i} - \frac{1}{M}\mathbf{1}
\right\|_2^2,
&
\mathcal{L}_{\mathrm{grp}}^{\mathrm{ent}}
&=
-\frac{1}{N}\sum_{i=1}^N H(P_\tau^{i}),
\end{align*}
where $\mathcal{L}_{\mathrm{grp}}^{\mathrm{bal}}$ encourages roughly balanced group usage and $\mathcal{L}_{\mathrm{grp}}^{\mathrm{ent}}$ encourages exploration early in training.
Finally, $\mathcal{L}_{\mathrm{grp}}^{\mathrm{edge}}$ encourages the learned affinity $G_{t_b}$ to place higher weight on agent pairs whose communication is predicted to be useful under our learned pairwise utility signal (Section~\ref{sec:scout_credit}).

\subsection{Three-headed policy and mailbox communication}
\label{sec:scout_policy_mailbox}

Each agent uses a shared policy $\pi_\theta(\cdot\mid u_t^i)$ with three heads: (i) an environment-action head, (ii) a binary send head $c_t^i\in\{0,1\}$, and (iii) a recipient head over $\rho_t^i\in\{1,\dots,N\}$. 
The backbone produces two descriptors (Section~\ref{sec:scout_repr}): $z_t^{\mathrm{msg},i}\in\mathbb{R}^d$, used as the agent's message representation, and $z_{t_b}^{\mathrm{grp},i}$, used for grouping at macro-step boundaries. 
When agent $i$ sends ($c_t^i=1$), it transmits the message vector
$x_t^i \;\triangleq\; z_t^{\mathrm{msg},i}\in\mathbb{R}^d$,
and samples a recipient $\rho_t^i$ from the recipient head (see the bottom panel of Figure~\ref{fig:scout_overview}).

\paragraph{Recipient bias from group affinity.}
Let $\phi_{t}^{\mathrm{recv},i}\in\mathbb{R}^N$ denote the recipient logits produced by agent $i$'s policy head at time $t$. 
To encourage within-group communication without hard constraints, we add a macro-step-dependent log-bias derived from the affinity matrix $G_{t_b}$:
\begin{equation*}
\tilde{\phi}_{t}^{\mathrm{recv},i}(j)
=
\phi_{t}^{\mathrm{recv},i}(j)
+\log\!\bigl(G_{t_b}^{ij}+\varepsilon\bigr),
\quad j\in\{1,\dots,N\},
\end{equation*}
where $t_b$ is the macro-step containing $t$ and $\varepsilon>0$ is a small constant. 
We then sample $\rho_t^i$ from $\mathrm{softmax}(\tilde{\phi}_{t}^{\mathrm{recv},i})$.

\paragraph{Mailbox aggregation (choice of $g$).}
As defined in Section~\ref{sec:problem_setting}, a recipient may receive a variable number of incoming messages; we therefore map the multiset of received message vectors to a fixed-dimensional mailbox $m_{t+1}^j$ using an aggregation operator $g(\cdot)$. This provides a permutation-invariant, fixed-size input to the policy at the next step. 
In our implementation we choose a simple parameter-free scaled dot-product attention rule \citep{vaswani2017attention}:
\begin{flalign}
m_{t+1}^j
=
\sum_{i:\,\rho_t^i=j} w_{ij}\, x_t^i,\quad
w_{ij}
=
\frac{\exp\!\bigl(\langle x_t^j, x_t^i\rangle/\sqrt{d}\bigr)}
{\sum\limits_{i':\,\rho_t^{i'}=j}\exp\!\bigl(\langle x_t^j, x_t^{i'}\rangle/\sqrt{d}\bigr)},
\label{eq:mailbox_attn}
\end{flalign}
where $d$ is the message dimension. Intuitively, the recipient's own message descriptor $x_t^j$ acts as a query that weights incoming message descriptors which act as keys and values.

\subsection{Group-aware critic and grouping advantage}
\label{sec:scout_critic}

\paragraph{Group-aware critic.}
Under CTDE, we learn a centralized value function that estimates expected discounted team return from the global state. To reduce output complexity for large populations, SCoUT predicts values at the \emph{group} level and converts them into per-agent value baselines, used for advantage estimation, using the soft assignments. Concretely, given $s_t$, the critic outputs
$\mathbf{v}_t=[v_t^1,\dots,v_t^M]^\top = V_\psi(s_t)\in\mathbb{R}^M$,
and the per-agent baseline is the assignment-weighted mixture
\begin{equation*}
V_t^i
=
\sum_{g=1}^M P_\tau^{i,g}\, v_t^g
=
\bigl(P_\tau \mathbf{v}_t\bigr)_i .
\end{equation*}
We train $V_\psi$ with the standard PPO/GAE value regression objective $\mathcal{L}_V$: we form bootstrapped targets from GAE \citep{schulman2016gae} and minimize a squared-error loss \citep{schulman2017ppo}.

\paragraph{Grouping advantage.}
At each macro-step boundary $t_b$, we derive an advantage signal for the grouping module by comparing the predicted per-agent value under the learned assignment to a permutation baseline. Let $\sigma$ be a random permutation of $\{1,\dots,M\}$ and define the permuted group-value vector $(\mathbf{v}_{t_b}^{\sigma})_g \triangleq \mathbf{v}_{t_b,\sigma(g)}$. We set
\[
A_{b}^{\mathrm{grp},i}
=
\underbrace{\bigl(P_\tau \mathbf{v}_{t_b}\bigr)_i}_{\text{value under learned grouping}}
-
\underbrace{\bigl(P_\tau \mathbf{v}_{t_b}^{\sigma}\bigr)_i}_{\text{permutation baseline}} .
\]
Subtracting the second term acts as a REINFORCE-style baseline. It preserves the scale of predicted values while being independent of the particular group labeling, and thus, reduces gradient variance for learning the grouping distribution \citep{williams1992reinforce,greensmith2004variance}.

\subsection{Communication critic and counterfactual mailbox credit assignment}
\label{sec:scout_credit}

A core difficulty in learned communication is credit assignment: multiple simultaneous messages can jointly influence future rewards.
SCoUT learns an auxiliary communication critic with two heads:
(i) a message-conditioned value head $V_{\omega}^{\mathrm{msg}}$ for recipient evaluation, and
(ii) a pairwise utility head $Q_{\omega}$ for sender--recipient utilities.
We train this critic by minimizing squared one-step Temporal Difference (TD) errors \citep{sutton1988learning} over a batch of $T$ primitive steps per update,
using Polyak averaging \citep{polyak1992acceleration,lillicrap2015continuous}.

\paragraph{Counterfactual send advantage.}
For a sender $i$ that addressed recipient $j=\rho_t^i$, we compute a counterfactual mailbox that removes $i$'s contribution while keeping all other incoming messages fixed, following the counterfactual credit assignment paradigm discussed in \cite{foerster2018coma}.
For a sender $i$ that addressed recipient $j=\rho_t^i$, let 
$m_{t+1}^j = g(\{x_t^{i'}:\, c_t^{i'}=1,\ \rho_t^{i'}=j\})$, as defined in Section~\ref{sec:problem_setting}.
We define the leave-one-out counterfactual mailbox by removing $i$'s message and re-applying the same aggregator $
m_{t+1}^{j\setminus i}
=
g\!\left(\{x_t^{i'}:\, c_t^{i'}=1,\ \rho_t^{i'}=j,\ i'\neq i\}\right).
$
In our implementation, where $g$ is the parameter-free attention rule in \eqref{eq:mailbox_attn}, this reduces to a normalized leave-one-out attention aggregate:
\[
m_{t+1}^{j\setminus i}
=
\sum_{i'\in \mathcal{S}_t(j)\setminus\{i\}}
\frac{w_{i'j}}{\sum_{k\in \mathcal{S}_t(j)\setminus\{i\}} w_{kj}}\; x_t^{i'}.
\]
Subsequently, we define the send advantage as
\begin{equation*}
A_{t}^{\mathrm{send},i}
=
V_{\omega}^{\mathrm{msg}}\!\bigl(s_{t+1},\, f_{t+1}^j(m_{t+1}^j)\bigr)
-
V_{\omega}^{\mathrm{msg}}\!\bigl(s_{t+1},\, f_{t+1}^j(m_{t+1}^{j\setminus i})\bigr),
\label{eq:send_cf_adv}
\end{equation*}
where $f_{t+1}^j(\cdot)$ denotes the recipient feature construction with the specified mailbox input.

\paragraph{Recipient advantage and edge-utility alignment.}
Let the pairwise utility matrix at time $t$ be
\[
U_t^{ij} := Q_{\omega}\!\bigl(s_{t+1},\, f_t^i,\, f_{t+1}^j\bigr),
\]
for valid sender-recipient pairs $(i,j)$.
For a sender $i$ with chosen recipient $\rho_t^i$, we form a centered recipient advantage
\[
A_t^{\mathrm{recv},i}
=
U_t^{i\rho_t^i}
-
\frac{1}{|\mathcal{R}_t^i|}\sum_{j\in \mathcal{R}_t^i} U_t^{ij},
\]
where $\mathcal{R}_t^i$ is the set of possible recipients for $i$.
To align grouping with communication utility, we define $\tilde{G}_{t_b}^{i,\cdot}$ and $\tilde{U}_{t_b}^{i,\cdot}$ as row-centered and $\ell_2$-normalized versions of the affinity and utility rows respectively, and set
\[
\mathcal{L}_{\mathrm{grp}}^{\mathrm{edge}}
=
-\frac{1}{N}\sum_{i=1}^N
\frac{\langle \tilde{G}_{t_b}^{i,\cdot},\, \tilde{U}_{t_b}^{i,\cdot}\rangle}
{\|\tilde{G}_{t_b}^{i,\cdot}\|\,\|\tilde{U}_{t_b}^{i,\cdot}\|}.
\]
This is akin to computing cosine similarity between row vectors of two matrices \citep{brockmeier2017similarity} and encourages groups to be \emph{communication-consistent}: agents are grouped together when the learned utility predicts that they are useful communication partners, and pushed apart otherwise.

\subsection{Overall optimization and decentralized execution}
\label{sec:scout_objective_exec}

We train the three-headed policy with PPO \citep{schulman2017ppo}, using (i) standard GAE advantages for environment actions \citep{schulman2016gae} and (ii) the communication advantages $A_t^{\mathrm{send},i}$ and $A_t^{\mathrm{recv},i}$ above for the send and recipient heads.
Let $\mathcal{L}_{\mathrm{PPO}}^{\mathrm{env}}, \mathcal{L}_{\mathrm{PPO}}^{\mathrm{send}}, \mathcal{L}_{\mathrm{PPO}}^{\mathrm{recv}}$ denote the standard clipped PPO surrogates for each head with their corresponding advantages.
The full objective combines policy, environment critic, communication critic, and grouping losses:
\begin{equation*}
\mathcal{L}
=
\mathcal{L}_{\mathrm{PPO}}^{\mathrm{env}}
+\mathcal{L}_{\mathrm{PPO}}^{\mathrm{send}}
+\mathcal{L}_{\mathrm{PPO}}^{\mathrm{recv}}
+c_V\mathcal{L}_V
+c_\omega\mathcal{L}_\omega
+c_{\mathrm{grp}}\mathcal{L}_{\mathrm{grp}},
\label{eq:total_loss}
\end{equation*}
where $\mathcal{L}_\omega$ denotes the communication critic loss described in Section~\ref{sec:scout_credit}. The coefficients $c_V,c_\omega,c_{\mathrm{grp}}\ge 0$ are user-chosen scalar weights that trade off the auxiliary losses against the PPO policy losses; we treat them as hyperparameters.
At test time, SCoUT discards centralized training components and executes only the shared decentralized policy per agent.

\section{Experiments}
\label{sec:experiments}

We evaluate \textsc{SCoUT} on two large-population multi-agent benchmarks with two goals:
(i) show that \textsc{SCoUT} can be trained directly at \emph{hundreds of agents} while retaining strong coordination; and
(ii) isolate which design choices drive these gains.

\subsection{Benchmarks}
\paragraph{MAgent Battle.}
We consider the \textsc{Battle} scenario in \textsc{MAgent}~\citep{zheng2018magent}, a competitive two-team gridworld where each team must coordinate to eliminate the opposing team under a fixed episode horizon. At each step, each agent chooses to either move or attack; attacks reduce the target's health, and agents are eliminated when their health reaches zero (additional details for \textsc{Battle} can be found in Appendix~\ref{sec:supp_battle}). We evaluate four population scales: \texttt{20\text{v}20}, \texttt{64\text{v}64}, \texttt{81\text{v}81}, and \texttt{100\text{v}100}. Notably, prior structured-communication results on this benchmark are typically reported up to \texttt{64\text{v}64}; our \texttt{81\text{v}81} and \texttt{100\text{v}100} settings explicitly stress-test scalability beyond that regime.

\paragraph{Pursuit (SISL).}
We also evaluate on \textsc{Pursuit} from the PettingZoo SISL suite~\citep{terry2021pettingzoo}, consisting of \texttt{P} pursuers and \texttt{E} evaders (\texttt{E}$<$\texttt{P}).
An evader is captured when all {free} adjacent cells around it are occupied by pursuers; thus the required number of pursuers is 4 in open space, 3 along walls/obstacles, and 2 in corners (additional details for \textsc{Pursuit} can be found in Appendix~\ref{sec:supp_pursuit}).
We use \textsc{Pursuit} primarily for targeted ablations to understand which components of \textsc{SCoUT} are necessary for strong coordination. We evaluate five scales: \texttt{20P-8E}, \texttt{40P-16E}, \texttt{60P-24E}, \texttt{80P-32E}, and \texttt{100P-40E}. Unlike the default Pursuit configuration (8 pursuers, 30 evaders)~\citep{terry2021pettingzoo,gerstgrasser2023selectively}, we study regimes with $E<P$ to stress-test scalability with respect to the number of learning agents while keeping capture milestones informative under a fixed 500-step horizon.

\noindent
Across all benchmarks, we use a fixed macro-step length $K{=}10$ and set the number of groups by a simple scale-dependent rule. For Battle: $M=\lfloor \sqrt{N}\rfloor$ with $N$ agents per team and for Pursuit: $M=\lfloor E/2\rfloor$ with $E$ evaders. Appendix~\ref{sec:hyperparameters} summarizes all the hyperparameters used in the experiments and Appendix~\ref{sec:sensitivity} presents a sensitivity study for the choice of $M$ and $K$.

\noindent\textbf{Compute.}
All experiments were run on a machine with an NVIDIA L40 GPU and an AMD CPU (32 cores @ 4.1\,GHz).

\subsection{Scalable competitive coordination on MAgent Battle}
\label{sec:battle_results}

\paragraph{Setup and claim under test.}
This experiment evaluates whether \textsc{SCoUT} maintains {robust} and {decisive} competitive coordination as population size grows.
Unlike previous work emphasizing zero-shot transfer from small-$N$ training, we train each method \emph{directly} at the target scale (up to \texttt{100v100}), stressing optimization stability in the large-instance regime. All methods are trained for 4M steps.

\paragraph{Baselines.}
We compare against one non-communication baseline and several learned-communication baselines:

\begin{enumerate}[label=(\roman*), leftmargin=*, itemsep=0pt, topsep=0pt, parsep=0pt, partopsep=0pt]
  \item \textbf{Independent deep Q-learning (IDQN)} ~\citep{rashid2020monotonic,mnih2015dqn}, where each agent learns from its local observation without explicit message exchange.


  \item \textbf{CommFormer}~\citep{hu2024learning} uses a transformer-style message encoder with attention over other agents’ communication tokens, enabling content-based routing and richer interaction modeling compared to fixed-topology schemes. We include it as a representative {attention-based} learned-communication baseline.

  \item \textbf{ExpoComm}~\citep{Li2025ExpoComm}, designed for many-agent communication via exponential-topology neighborhoods. We evaluate \emph{Peer-$n{=}7$} (time-varying neighborhood) and \emph{Static-$n{=}7$} (fixed neighborhood) under the same per-agent recipient budget.
\end{enumerate}
\vspace{-2mm}

\begin{figure}[t]
  \centering
  \includegraphics[width=0.85\linewidth]{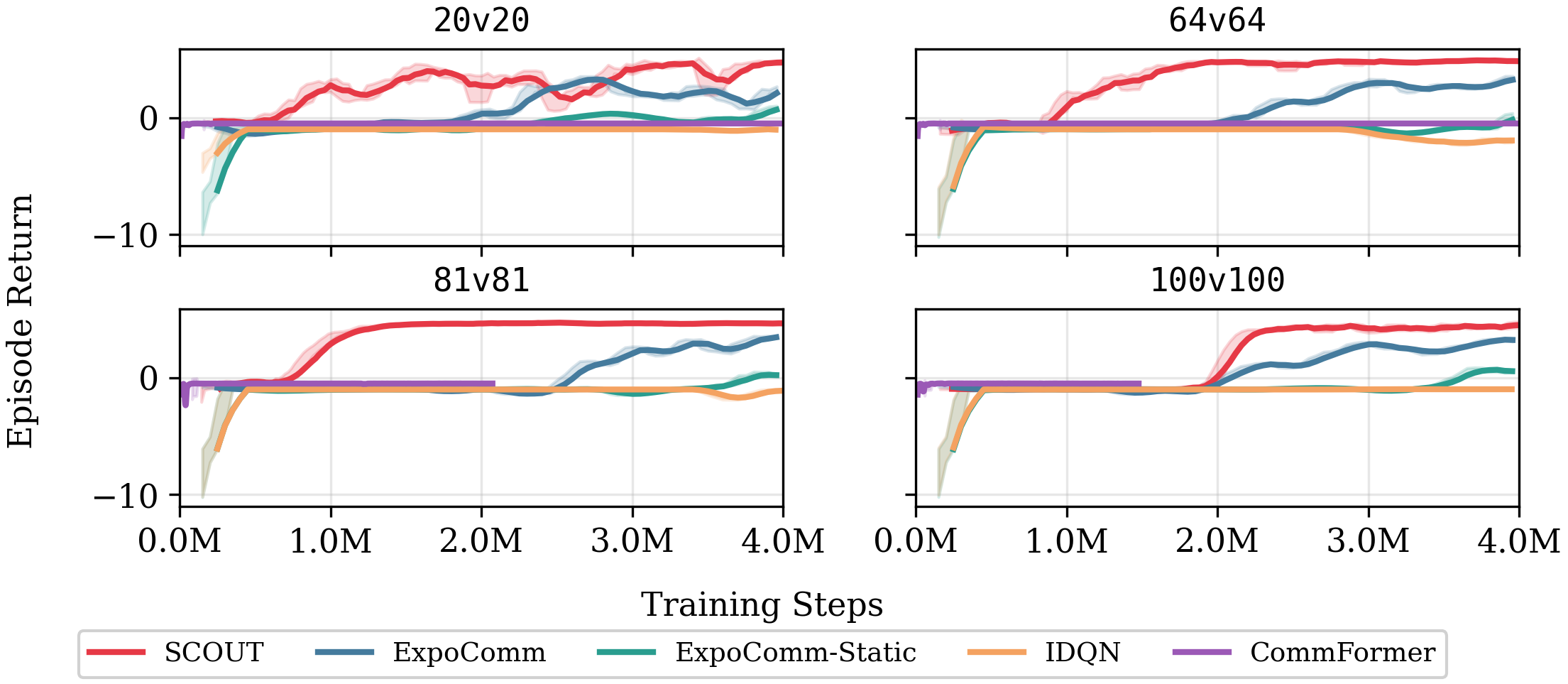}
  \caption{\textbf{Battle training curves across scales.}
  Episode return vs environment steps (shaded regions indicate variability across seeds).
  \textsc{SCoUT} learns rapidly and consistently across scales, while baselines exhibit substantial scale- and topology-sensitivity.}
  \label{fig:battle_training_curves}
  \vspace{-2mm}
\end{figure}

\paragraph{Metrics and evaluation protocol.}
We evaluate each trained policy on 20 independent evaluation seeds per setting.
We report:
(i) \textbf{win rate} (\% episodes where the red team finishes with more agents alive than the blue team);
(ii) \textbf{elimination rate} (\% blue eliminated, i.e., blue kills divided by $N$); and
(iii) \textbf{decisiveness} via time-to-milestone:
$\mathrm{TT}_{50}$ / $\mathrm{TT}_{75}$ = steps to reach 50\% / 75\% blue elimination (horizon=200).
We additionally report \textbf{reach rates} R$_{50}$/R$_{75}$, the \% of evaluation episodes that reach each milestone within the horizon.
We report $\mathrm{TT}_k$ only when R$_k \ge 50\%$ (otherwise, \texttt{N/A}).
Furthermore, we plot \textbf{training curves} (episode return vs.\ environment steps; Figure~\ref{fig:battle_training_curves}) to assess optimization stability and sample efficiency when training \emph{directly} at each target scale. Finally, to characterize how learned communication aligns with the discovered interaction structure, we report the in-group message fraction across all scales in Appendix~\ref{sec:ingroup-frac}(Table~\ref{tab:ingroup_all_scales}). We also present qualitative rollouts for the \texttt{100v100} scenario (Appendix~\ref{sec:qual_battle}).

\begin{table}[t]
\centering
\scriptsize
\setlength{\tabcolsep}{2.6pt}
\renewcommand{\arraystretch}{1.15}
\newcommand{\unc}[1]{{\tiny$\pm$#1}}
\resizebox{\linewidth}{!}{%
\begin{tabular}{lcccc|cccc}
\toprule
\rowcolor{tablegray}
\textbf{Method} & \textbf{20v20} & \textbf{64v64} & \textbf{81v81} & \textbf{100v100} &
\textbf{20v20} & \textbf{64v64} & \textbf{81v81} & \textbf{100v100} \\
\midrule

\multicolumn{5}{l}{\textit{Win rate (\%)}} &
\multicolumn{4}{l}{\textit{Elimination rate (\%)} (blue eliminated / $N$)}\\
\textsc{SCoUT} & \best{100.0\unc{0.0}} & \best{100.0\unc{0.0}} & \best{100.0\unc{0.0}} & \best{100.0\unc{0.0}} &
\best{95\unc{6}} & \best{98\unc{4}} & \best{99\unc{3}} & \best{99\unc{3}}\\
ExpoComm & 0.0\unc{0.0} & 95.0\unc{4.9} & 96.0\unc{4.2} & 96.4\unc{2.7} &
11\unc{16} & 94\unc{12} & 72\unc{16} & 75\unc{21}\\
ExpoComm-Static & 0.0\unc{0.0} & 0.0\unc{0.0} & 60.0\unc{11.0} & 65.0\unc{10.7} &
3\unc{8} & 1\unc{2} & 57\unc{30} & 36\unc{10}\\
IDQN & 0.0\unc{0.0} & 0.0\unc{0.0} & 0.0\unc{0.0} & 0.0\unc{0.0} &
0 & 0 & 7\unc{2} & 0\\
CommFormer  & 0.0\unc{0.0} & 0.0\unc{0.0} & 0.0\unc{0.0} & 0.0\unc{0.0} &
0 & 0 & 0 & 0\\
\midrule

\multicolumn{5}{l}{\textit{Milestone reach rate (\%) (R$_{50}$ / R$_{75}$)}} &
\multicolumn{4}{l}{\textit{$\mathrm{TT}_{50}$ / $\mathrm{TT}_{75}$ (steps; horizon=200)}}\\
\textsc{SCoUT} & \best{100 / 100} & \best{100 / 100} & \best{100 / 100} & \best{100 / 100} &
\best{24\unc{1} / 30\unc{1}} & \best{24\unc{2} / 29\unc{1}} & \best{27\unc{3} / 32\unc{2}} & \best{31\unc{4} / 39\unc{2}}\\
ExpoComm & 0 / 0 & 95 / 95 & 90 / 55 & 80 / 60 &
\texttt{N/A} & 59\unc{20} / 81\unc{24} & 100\unc{44} / 182\unc{14} & 101\unc{21} / 160\unc{23}\\
ExpoComm-Static & 0 / 0 & 0 / 0 & 60 / 60 & 10 / 0 &
\texttt{N/A} & \texttt{N/A} & 85\unc{7} / 153\unc{20} & \texttt{N/A}\\
IDQN & 0 / 0 & 0 / 0 & 0 / 0 & 0 / 0 &
\texttt{N/A} & \texttt{N/A} & \texttt{N/A} & \texttt{N/A}\\
CommFormer &0 / 0 & 0 / 0 & 0 / 0 & 0 / 0 &
 \texttt{N/A} & \texttt{N/A} & \texttt{N/A} & \texttt{N/A}\\
\bottomrule
\end{tabular}%
}
\caption{\textbf{MAgent Battle results across scales (20 evaluation seeds).}
We report win rate, elimination rate, \textbf{milestone reach rates} R$_{50}$/R$_{75}$, and \textbf{decisiveness} via $\mathrm{TT}_{50}/\mathrm{TT}_{75}$.}
\label{tab:battle_outcomes}
\vspace{-5mm}
\end{table}

\paragraph{Results and discussion.}
Table~\ref{tab:battle_outcomes} and Figure~\ref{fig:battle_training_curves} show two consistent patterns\footnote{CommFormer training curves are truncated for larger scales: \texttt{81v81} reached $\approx$50\% of the planned 4M steps while \texttt{100v100} reached $\approx$35\% (both after 6 days, $\sim$4$\times$ the wall-clock budget used for \textsc{SCoUT}).}.
\textbf{(1) \textsc{SCoUT} trains reliably across all scales and is highly repeatable.}
\textsc{SCoUT} achieves 100\% win rate for all $N$ and sustains near-complete elimination (95--99\%), with elimination and milestone times exhibiting very low variance across 20 evaluation seeds.
Training is similarly stable: across all four scales, \textsc{SCoUT} shows a clear improvement phase followed by stable convergence with substantially less across-seed dispersion than the baselines (Figure~\ref{fig:battle_training_curves}); notably, this stability persists even at the largest populations (\texttt{81v81} and \texttt{100v100}), where \textsc{SCoUT} converges smoothly and quickly compared to the baselines.
At the other end of the spectrum (\texttt{20v20}), \textsc{SCoUT} still learns reliably and significantly outperforms the baselines, while \textsc{ExpoComm} is consistently poor and shows no sustained learning progress.
Across scales, the static-topology variant underperforms its adaptive counterpart, suggesting that fixed neighborhoods are overly restrictive in dynamic combat where effective coordination partners shift over time.
\textbf{(2) \textsc{SCoUT} converts wins into decisive elimination more consistently and more quickly.}
At larger $N$, ExpoComm baselines can achieve high win rate, but they often leave substantial opponent mass and fail to reach 75\% elimination within the horizon; baseline milestone reach rates degrade with scale, indicating that many winning episodes do not consistently translate advantage into high-elimination outcomes.
Even when milestones are reached, \textsc{SCoUT} attains them substantially faster (Table~\ref{tab:battle_outcomes}), consistent with its smoother and more stable learning dynamics (Figure~\ref{fig:battle_training_curves}). Notably, it is also the only method with R$_{50}{=}\text{R}_{75}{=}100\%$ at every scale, confirming consistently decisive elimination rather than occasional fast wins.

\subsection{Ablations on Pursuit}
\label{sec:pursuit_ablations}

\paragraph{Setup and claim under test.}
We use \textsc{Pursuit} to isolate which ingredients of \textsc{SCoUT} drive scalable multi-agent coordination in cooperative capture.
All methods are trained {directly} at each target scale for 4M steps.

\paragraph{Variants and baseline.}
We compare the full method against targeted removals and a scalable-communication baseline:
\begin{enumerate}[label=(\roman*), leftmargin=*, itemsep=0pt, topsep=0pt, parsep=0pt, partopsep=0pt]
    \item \textbf{\textsc{SCoUT}:} learned grouping + group-aware critic + counterfactual communication advantages.
    \item \textbf{w/o counterfactual:} removes the counterfactual decomposition for communication decisions (uses standard Generalized Advantage Estimation \citep{schulman2016gae}).
    \item \textbf{w/o grouping:} removes learned grouping; consequently, the group-aware critic is disabled.
    \item \textbf{ExpoComm (Peer-$n{=}7$):} same learned communication baseline as Section~\ref{sec:battle_results}.
\end{enumerate}
\vspace{-2mm}

\begin{figure}[t]
  \centering
  \includegraphics[width=0.75\linewidth]{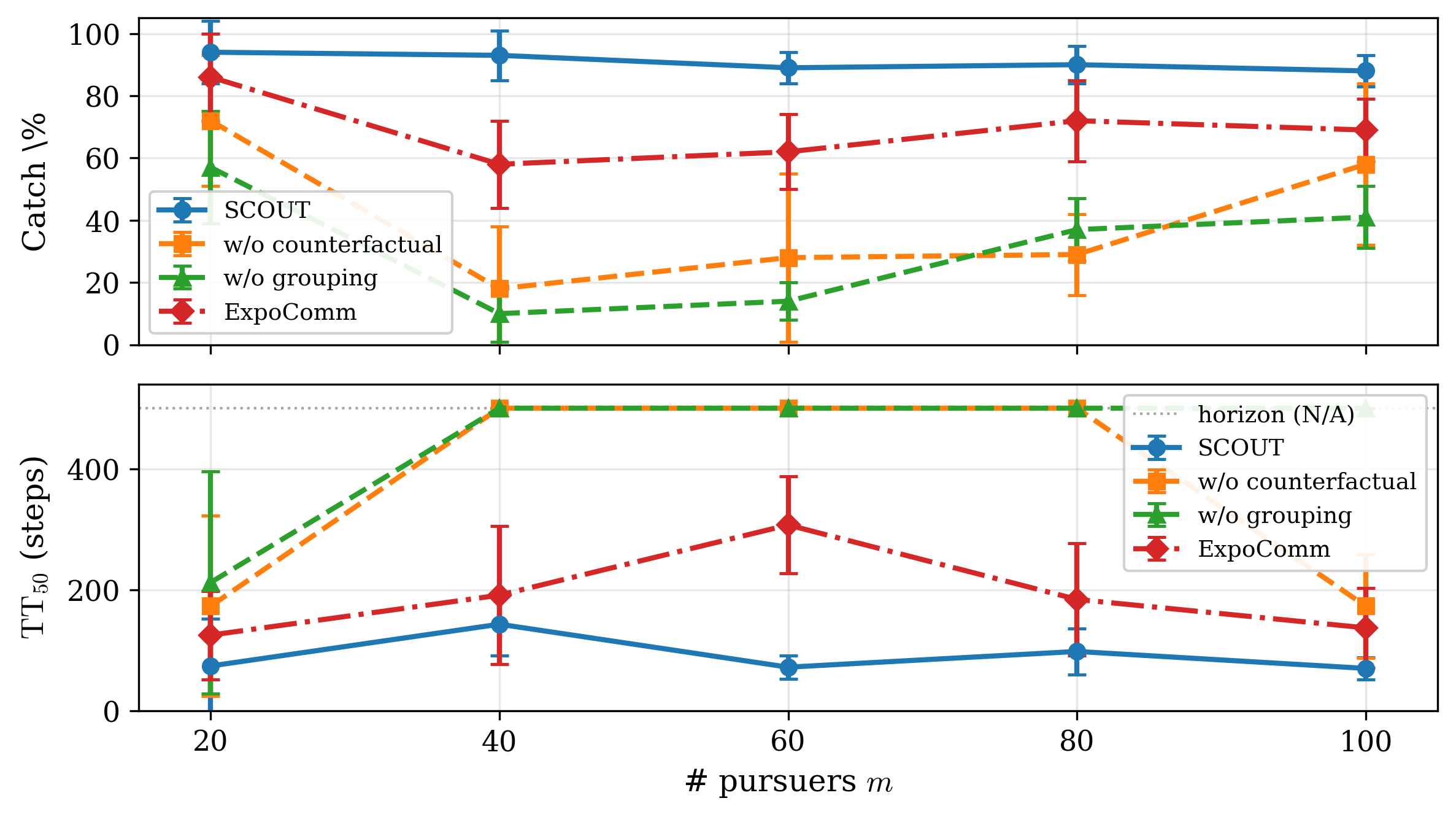}
  \caption{\textbf{Pursuit scaling summary.}
  Catch\% (top) and $\mathrm{TT}_{50}$ (bottom) vs.\ number of pursuers $m$ (error bars: std over 20 evaluation seeds).
  In the bottom panel, we plot $\mathrm{TT}_{50}{=}500$ when R$_{50}{<}50\%$ (Table~\ref{tab:pursuit_ablations} reports these cases as \texttt{N/A}).}
  \label{fig:pursuit_scaling}
  \vspace{-2mm}
\end{figure}

\paragraph{Metrics and evaluation protocol.}
We evaluate each trained policy on 20 independent evaluation seeds per setting.
We report:
(i) \textbf{Catch\%} (fraction of evaders captured);
(ii) \textbf{Done\%} (\% episodes capturing all evaders); and
(iii) \textbf{decisiveness} using the same milestone metrics as in Section~\ref{sec:battle_results}:
R$_{50}$/R$_{75}$ (milestone reach rate) and $\mathrm{TT}_{50}/\mathrm{TT}_{75}$ (time-to-milestone),
but with milestones defined over the \emph{capture fraction} rather than elimination.
Furthermore, we visualize scaling trends with Catch\% and $\mathrm{TT}_{50}$ vs.\ population size (Figure~\ref{fig:pursuit_scaling}). Finally, similar to battle, we also report the in-group message fraction for \textsc{Pursuit} in Appendix~\ref{sec:ingroup-frac} (Table~\ref{tab:ingroup_all_scales}) along with qualitative rollouts for \texttt{100P-40E} (Appendix~\ref{sec:qual_pursuit}).

\begin{table}[t]
\centering
\scriptsize
\setlength{\tabcolsep}{2.6pt}
\renewcommand{\arraystretch}{1.15}
\newcommand{\unc}[1]{{\tiny$\pm$#1}}

\resizebox{\linewidth}{!}{%
\begin{tabular}{lccccc|ccccc}
\toprule
\rowcolor{tablegray}
\textbf{Method} & \textbf{20P-8E} & \textbf{40P-16E} & \textbf{60P-24E} & \textbf{80P-32E} & \textbf{100P-40E} &
\textbf{20P-8E} & \textbf{40P-16E} & \textbf{60P-24E} & \textbf{80P-32E} & \textbf{100P-40E} \\
\midrule
\multicolumn{6}{l}{\textit{Catch rate (Catch\%)}} &
\multicolumn{5}{l}{\textit{Completion rate (Done\%)}}\\
\textsc{SCoUT} & \best{94\unc{10}} & \best{93\unc{8}} & \best{89\unc{5}} & \best{90\unc{6}} & \best{88\unc{5}} &
\best{70} & \best{50} & \best{0} & \best{5} & \best{0}\\
w/o counterfactual & 72\unc{21} & 18\unc{20} & 28\unc{27} & 29\unc{13} & 58\unc{26} &
20 & 0 & 0 & 0 & 0\\
w/o grouping & 57\unc{18} & 10\unc{9} & 14\unc{6} & 37\unc{10} & 41\unc{10} &
0 & 0 & 0 & 0 & 0\\
ExpoComm (Peer-$n{=}7$) & 86\unc{14} & 58\unc{14} & 62\unc{12} & 72\unc{13} & 69\unc{10} &
40 & 0 & 0 & 0 & 0\\
\bottomrule
\end{tabular}%
}

\vspace{1.5mm}

\resizebox{\linewidth}{!}{%
\begin{tabular}{lccccc}
\toprule
\rowcolor{tablegray}
\textbf{Method} & \textbf{20P-8E} & \textbf{40P-16E} & \textbf{60P-24E} & \textbf{80P-32E} & \textbf{100P-40E} \\
\midrule
\multicolumn{6}{l}{\textit{$\mathrm{TT}_{50}$ (steps; horizon=500) \; (R$_{50}$\% in parentheses)}}\\
\textsc{SCoUT} & \best{74\unc{78} (100)} & \best{143\unc{52} (100)} & \best{72\unc{19} (100)} & \best{98\unc{38} (100)} & \best{70\unc{18} (100)}\\
w/o counterfactual & 173\unc{149} (90) & \texttt{N/A} (15) & \texttt{N/A} (20) & \texttt{N/A} (5) & 173\unc{86} (55)\\
w/o grouping & 212\unc{184} (80) & \texttt{N/A} (0) & \texttt{N/A} (0) & \texttt{N/A} (15) & \texttt{N/A} (20)\\
ExpoComm (Peer-$n{=}7$) & 125\unc{73} (100) & 191\unc{114} (80) & 307\unc{80} (85) & 184\unc{93} (100) & 137\unc{66} (95)\\
\midrule
\multicolumn{6}{l}{\textit{$\mathrm{TT}_{75}$ (steps; horizon=500) \; (R$_{75}$\% in parentheses)}}\\
\textsc{SCoUT} & \best{138\unc{82} (95)} & \best{227\unc{65} (95)} & \best{139\unc{64} (100)} & \best{212\unc{79} (100)} & \best{149\unc{41} (100)}\\
w/o counterfactual & 214\unc{136} (65) & \texttt{N/A} (0) & \texttt{N/A} (15) & \texttt{N/A} (0) & \texttt{N/A} (35)\\
w/o grouping & \texttt{N/A} (30) & \texttt{N/A} (0) & \texttt{N/A} (0) & \texttt{N/A} (0) & \texttt{N/A} (0)\\
ExpoComm (Peer-$n{=}7$) & 234\unc{114} (80) & \texttt{N/A} (15) & \texttt{N/A} (20) & \texttt{N/A} (40) & 425\unc{59} (50)\\
\bottomrule
\end{tabular}%
}

\caption{\textbf{Pursuit ablations across scales (20 evaluation seeds; horizon=500).}
We report Catch\%, Done\%, and milestone times $\mathrm{TT}_{50}/\mathrm{TT}_{75}$.
For milestone rows, each entry is shown as $\mathrm{TT}_k$ (R$_k$), where R$_k$ is the \% of episodes reaching the $k$\% capture milestone.}
\label{tab:pursuit_ablations}
\vspace{-2mm}
\end{table}

\paragraph{Results and discussion.}
Table~\ref{tab:pursuit_ablations} and Figure~\ref{fig:pursuit_scaling} reveal two consistent patterns.
\textbf{(1) Full \textsc{SCoUT} scales with high capture and reliable milestone reach.}
Across all five scales, \textsc{SCoUT} maintains high Catch\% with R$_{50}{=}100\%$, meaning every evaluation episode reaches 50\% capture within 500 steps.
It also reaches 75\% capture reliably at moderate and large scales, with consistently small milestone times and stable across-seed error bars (Figure~\ref{fig:pursuit_scaling}).
We report Done\% as full capture within the fixed 500-step horizon; Done\% naturally decreases with scale (often to 0\%) because the number of evaders increases while full capture requires simultaneously surrounding \emph{every} remaining evader, making complete capture within 500 steps increasingly unlikely even when partial coordinated capture is strong. \textbf{(2) Isolation experiments show that counterfactual credit and temporal grouping drive scalability.}
Removing counterfactual advantages causes an abrupt scaling failure: beyond \texttt{20P-8E}, Catch\% collapses and becomes highly variable, and milestone reach drops sharply--R$_{50}$ falls to 15-20\% for \texttt{40P-16E}/\texttt{60P-24E} and 5\% for \texttt{80P-32E}), making $\mathrm{TT}_{50}$/$\mathrm{TT}_{75}$ undefined. At \texttt{100P-40E} it only partially recovers to R$_{50}{=}55\%$ while still failing to reach 75\% reliably.
Removing grouping is similarly brittle at scale: Catch\% drops to 10-41\% for \texttt{40P-16E} through \texttt{100P-40E}, with R$_{50}\le 20\%$ and R$_{75}{=}0\%$ throughout these larger scenarios, indicating persistent failure to form effective capture coalitions.
Together, these ablations support the conclusion that \textsc{SCoUT}'s counterfactual communication learning and temporal grouping are both necessary to sustain high capture and early milestone attainment as the population grows.

\section{Conclusion}
\label{sec:conclusion}

Scaling learned communication to large teams is difficult because sender-receiver interactions grow combinatorially and message-level credit becomes noisy. We presented \textsc{SCoUT}, an approach which makes communication scalable by learning a slowly-varying latent interaction structure: it resamples soft groups at a macro-step timescale, uses the resulting affinities as a differentiable prior over recipients, and reuses the same structure in a group-aware critic for tractable value learning at large $N$. \textsc{SCoUT} further introduces a counterfactual mailbox mechanism that isolates each sender’s marginal contribution to a recipient’s aggregated message, yielding fine-grained learning signals for both sending and recipient selection. Across large-population benchmarks, \textsc{SCoUT} trains directly at target scale and maintains strong, reliable coordination as $N$ grows. In \textsc{MAgent Battle}, it achieves perfect win rate with near-complete elimination and consistently early milestones, while state-of-the-art baselines are less decisive and less reliable. In \textsc{Pursuit}, \textsc{SCoUT} sustains high capture with early milestones across scales. Ablations confirm that temporal grouping and counterfactual communication credit are both necessary: removing either causes sharp drops in performance at moderate and large scales.


\paragraph{Limitations and future work.}
\textsc{SCoUT} adds training-time structure, in particular the macro-step $K$ and number of groups $M$, which we currently set as fixed hyperparameters. Although one setting worked well across our benchmarks, performance can depend on how well $K$ matches the interaction timescale and $M$ matches agent behavior heterogeneity. In addition, our leave-one-out mailbox counterfactual provides clean marginal credit for senders under a fixed aggregation rule, but does not capture multi-message interaction effects. Future work includes adapting $K$ and $M$ online using signals like recipient entropy or utility drift, extending credit assignment to model message interactions, and evaluating \textsc{SCoUT} under continuous control, heterogeneity, and explicit communication/energy budgets.

\newpage

\bibliography{main}

@inproceedings{Singh2019IC3Net,
  author    = {Amanpreet Singh and Tushar Jain and Sainbayar Sukhbaatar},
  title     = {Learning when to Communicate at Scale in Multiagent Cooperative and Competitive Tasks},
  booktitle = {7th International Conference on Learning Representations (ICLR)},
  year      = {2019}
}

@inproceedings{Das2019TarMAC,
  author    = {Abhishek Das and Th{\'{e}}ophile Gervet and Joshua Romoff and Dhruv Batra and Devi Parikh and Mike Rabbat and Joelle Pineau},
  title     = {{TarMAC}: Targeted Multi-Agent Communication},
  booktitle = {36th International Conference on Machine Learning (ICML)},
  pages     = {1538--1546},
  year      = {2019}
}

@inproceedings{Liu2020G2ANet,
  author    = {Yong Liu and Weixun Wang and Yujing Hu and Jianye Hao and Xingguo Chen and Yang Gao},
  title     = {Multi-Agent Game Abstraction via Graph Attention Neural Network},
  booktitle = {34th AAAI Conference on Artificial Intelligence},
  year      = {2020},
  pages     = {7211--7218}
}

@inproceedings{Li2025ExpoComm,
  author    = {Xinran Li and Xiaolu Wang and Chenjia Bai and Jun Zhang},
  title     = {Exponential Topology-enabled Scalable Communication in Multi-agent Reinforcement Learning},
  booktitle = {13th International Conference on Learning Representations (ICLR)},
  year      = {2025}
}

@inproceedings{Zheng2018MAgent,
  title={Magent: A many-agent reinforcement learning platform for artificial collective intelligence},
  author={Zheng, Lianmin and Yang, Jiacheng and Cai, Han and Zhou, Ming and Zhang, Weinan and Wang, Jun and Yu, Yong},
  booktitle={32nd AAAI Conference on Artificial Intelligence},
  year={2018}
}

@inproceedings{bernstein2002complexity,
  title     = {The Complexity of Decentralized Control of {M}arkov Decision Processes},
  author    = {Bernstein, Daniel S. and Givan, Robert and Immerman, Neil and Zilberstein, Shlomo},
  booktitle = {18th Conference on Uncertainty in Artificial Intelligence (UAI)},
  year      = {2002}
}

@inproceedings{littman1994markov,
  title     = {Markov Games as a Framework for Multi-Agent Reinforcement Learning},
  author    = {Littman, Michael L.},
  booktitle = {11th International Conference on Machine Learning (ICML)},
  year      = {1994}
}

@article{busoniu2008survey,
  title   = {A Comprehensive Survey of Multiagent Reinforcement Learning},
  author  = {Bu{\c{s}}oniu, Lucian and Babu{\v{s}}ka, Robert and De Schutter, Bart},
  journal = {IEEE Transactions on Systems, Man, and Cybernetics, Part C (Applications and Reviews)},
  volume  = {38},
  number  = {2},
  pages   = {156--172},
  year    = {2008}
}

@article{hernandezleal2019survey,
author = {Hernandez-Leal, Pablo and Kartal, Bilal and Taylor, Matthew E.},
title = {A Survey and Critique of Multiagent Deep Reinforcement Learning},
year = {2019},
volume = {33},
number = {6},
journal = {Autonomous Agents and Multi-Agent Systems},
pages = {750–797}
}

@inproceedings{sukhbaatar2016commnet,
author = {Sukhbaatar, Sainbayar and Szlam, Arthur and Fergus, Rob},
title = {Learning Multiagent Communication with Backpropagation},
year = {2016},
booktitle = {30th International Conference on Neural Information Processing Systems (NIPS)},
pages = {2252–2260}
}

@inproceedings{foerster2016dial,
author = {Foerster, Jakob N. and Assael, Yannis M. and de Freitas, Nando and Whiteson, Shimon},
title = {Learning to Communicate with Deep Multi-Agent Reinforcement Learning},
year = {2016},
booktitle = {30th International Conference on Neural Information Processing Systems (NIPS)},
pages = {2145–2153}
}

@inproceedings{kim2019schednet,
  title         = {Learning to Schedule Communication in Multi-agent Reinforcement Learning},
  author        = {Kim, Daewoo and Moon, Sangwoo and Hostallero, David and Kang, Wan Ju and Lee, Taeyoung and Son, Kyunghwan and Yi, Yung},
  booktitle       = {7th International Conference on Learning Representations (ICLR)},
  year          = {2019}
}

@inproceedings{jiang2018atoc,
author = {Jiang, Jiechuan and Lu, Zongqing},
title = {Learning Attentional Communication for Multi-Agent Cooperation},
year = {2018},
booktitle = {32nd International Conference on Neural Information Processing Systems (NIPS)},
pages = {7265–7275}
}

@inproceedings{wang2021tom2c,
  title         = {{ToM2C}: Target-oriented Multi-agent Communication and Cooperation with Theory of Mind},
  author        = {Wang, Yuanfei and Zhong, Fangwei and Xu, Jing and Wang, Yizhou},
  booktitle       = {10th International Conference on Learning Representations (ICLR)},
  year          = {2022}
}

@inproceedings{foerster2018coma,
author = {Foerster, Jakob N. and Farquhar, Gregory and Afouras, Triantafyllos and Nardelli, Nantas and Whiteson, Shimon},
title = {Counterfactual Multi-Agent Policy Gradients},
year = {2018},
booktitle = {32nd AAAI Conference on Artificial Intelligence and Thirtieth Innovative Applications of Artificial Intelligence Conference and Eighth AAAI Symposium on Educational Advances in Artificial Intelligence}
}

@InProceedings{yang2018meanfield,
  title = 	 {Mean Field Multi-Agent Reinforcement Learning},
  author =       {Yang, Yaodong and Luo, Rui and Li, Minne and Zhou, Ming and Zhang, Weinan and Wang, Jun},
  booktitle = 	 {35th International Conference on Machine Learning (ICML)},
  pages = 	 {5571--5580},
  year = 	 {2018}
}

@article{schulman2017ppo,
  title         = {Proximal Policy Optimization Algorithms},
  author        = {Schulman, John and Wolski, Filip and Dhariwal, Prafulla and Radford, Alec and Klimov, Oleg},
  journal       = {arXiv preprint arXiv:1707.06347},
  year          = {2017},
  eprint        = {1707.06347},
  archivePrefix = {arXiv},
  primaryClass  = {cs.LG}
}

@inproceedings{schulman2016gae,
title = {High-Dimensional Continuous Control Using Generalized Advantage Estimation},
author = {John Schulman and Philipp Moritz and Sergey Levine and Michael Jordan and Pieter Abbeel},
booktitle = {4th International Conference on Learning Representations (ICLR)},
year  = 2016
}

@inproceedings{jang2017gumbel,
  title         = {Categorical Reparameterization with {G}umbel-Softmax},
  author        = {Jang, Eric and Gu, Shixiang and Poole, Ben},
  booktitle       = {5th International Conference on Learning Representations ({ICLR})},
  year          = {2017}
}

@inproceedings{maddison2017concrete,
  title         = {The Concrete Distribution: A Continuous Relaxation of Discrete Random Variables},
  author        = {Maddison, Chris J. and Mnih, Andriy and Teh, Yee Whye},
  booktitle       = {5th International Conference on Learning Representations ({ICLR})},
  year          = {2017}
}

@inproceedings{lowe2017maddpg,
author = {Lowe, Ryan and Wu, Yi and Tamar, Aviv and Harb, Jean and Abbeel, Pieter and Mordatch, Igor},
title = {Multi-Agent Actor-Critic for Mixed Cooperative-Competitive Environments},
year = {2017},
booktitle = {31st International Conference on Neural Information Processing Systems (NIPS)},
pages = {6382–6393}
}

@inproceedings{sunehag2018vdn,
author = {Sunehag, Peter and Lever, Guy and Gruslys, Audrunas and Czarnecki, Wojciech Marian and Zambaldi, Vinicius and Jaderberg, Max and Lanctot, Marc and Sonnerat, Nicolas and Leibo, Joel Z. and Tuyls, Karl and Graepel, Thore},
title = {Value-Decomposition Networks For Cooperative Multi-Agent Learning Based On Team Reward},
year = {2018},
booktitle = {17th International Conference on Autonomous Agents and MultiAgent Systems (AAMAS)}
}

@inproceedings{son2019qtran,
  title         = {{QTRAN}: Learning to Factorize with Transformation for Cooperative Multi-Agent Reinforcement Learning},
  author        = {Son, Kyunghwan and Kim, Daewoo and Kang, Wonseok and Hostallero, David and Yi, Yung},
  booktitle={36th International Conference on Machine Learning (ICML)},
  pages={5887--5896},
  year          = {2019}
}

@inproceedings{wang2021qplex,
  title         = {{QPLEX}: Duplex Dueling Multi-Agent {Q}-Learning},
  author        = {Wang, Jianhao and Ren, Zhizhou and Liu, Terry and Yu, Yang and Zhang, Chongjie},
  booktitle       = {9th International Conference on Learning Representations ({ICLR})},
  year          = {2021}
}

@inproceedings{yu2021mappo,
author = {Yu, Chao and Velu, Akash and Vinitsky, Eugene and Gao, Jiaxuan and Wang, Yu and Bayen, Alexandre and Wu, Yi},
title = {The Surprising Effectiveness of {PPO} in Cooperative, Multi-Agent Games},
year = {2022},
booktitle = {36th International Conference on Neural Information Processing Systems (NIPS)}
}

@inproceedings{samvelyan2019smac,
author = {Samvelyan, Mikayel and Rashid, Tabish and Schroeder de Witt, Christian and Farquhar, Gregory and Nardelli, Nantas and Rudner, Tim G. J. and Hung, Chia-Man and Torr, Philip H. S. and Foerster, Jakob and Whiteson, Shimon},
title = {The StarCraft Multi-Agent Challenge},
year = {2019},
booktitle = {18th International Conference on Autonomous Agents and MultiAgent Systems (AAMAS)},
pages = {2186–2188}
}

@inproceedings{christianos2021scaling,
  title={Scaling multi-agent reinforcement learning with selective parameter sharing},
  author={Christianos, Filippos and Papoudakis, Georgios and Rahman, Muhammad A. and Albrecht, Stefano V.},
  booktitle={38th International Conference on Machine Learning (ICML)},
  pages={1989--1998},
  year={2021}
}

@article{terry2020revisiting,
  title={Revisiting parameter sharing in multi-agent deep reinforcement learning},
  author={Terry, Justin K and Grammel, Nathaniel and Son, Sanghyun and Black, Benjamin and Agrawal, Aakriti},
  journal={arXiv preprint arXiv:2005.13625},
  year={2020}
}

@article{williams1992reinforce,
  title        = {Simple Statistical Gradient-Following Algorithms for Connectionist Reinforcement Learning},
  author       = {Williams, Ronald J.},
  journal      = {Machine Learning},
  volume       = {8},
  number       = {3--4},
  pages        = {229--256},
  year         = {1992},
  publisher    = {Springer}
}

@inproceedings{cho2014learning,
  title        = {Learning Phrase Representations using {RNN} Encoder--Decoder for Statistical Machine Translation},
  author       = {Cho, Kyunghyun and van Merri{\"e}nboer, Bart and G{\"u}l{\c{c}}ehre, {\c{C}}a{\u{g}}lar and Bahdanau, Dzmitry and Bougares, Fethi and Schwenk, Holger and Bengio, Yoshua},
  booktitle    = {Conference on Empirical Methods in Natural Language Processing (EMNLP)},
  pages        = {1724--1734},
  year         = {2014}
}

@inproceedings{vaswani2017attention,
  title     = {Attention Is All You Need},
  author    = {Vaswani, Ashish and Shazeer, Noam and Parmar, Niki and Uszkoreit, Jakob and Jones, Llion and Gomez, Aidan N. and Kaiser, {\L}ukasz and Polosukhin, Illia},
  booktitle = {31st International Conference on Neural Information Processing Systems},
  year      = {2017}
}

@inproceedings{lillicrap2015continuous,
  title   = {Continuous Control with Deep Reinforcement Learning},
  author  = {Lillicrap, Timothy P. and Hunt, Jonathan J. and Pritzel, Alexander and Heess, Nicolas and Erez, Tom and Tassa, Yuval and Silver, David and Wierstra, Daan},
  booktitle = {4th International Conference on Learning Represenations (ICLR)},
  year    = {2016}
}

@article{polyak1992acceleration,
  title   = {Acceleration of Stochastic Approximation by Averaging},
  author  = {Polyak, Boris T. and Juditsky, Anatoli B.},
  journal = {SIAM Journal on Control and Optimization},
  volume  = {30},
  number  = {4},
  pages   = {838--855},
  year    = {1992},
  publisher = {SIAM}
}

@article{greensmith2004variance,
  title={Variance reduction techniques for gradient estimates in reinforcement learning},
  author={Greensmith, Evan and Bartlett, Peter L and Baxter, Jonathan},
  journal={Journal of Machine Learning Research},
  volume={5},
  pages={1471--1530},
  year={2004}
}

@article{brockmeier2017similarity,
  title   = {Quantifying the Informativeness of Similarity Measurements},
  author  = {Brockmeier, Austin J. and others},
  journal = {Journal of Machine Learning Research},
  year    = {2017},
  volume  = {18},
  number  = {1},
  pages   = {1--41}
}

@inproceedings{niu2021multi,
  title={Multi-Agent Graph-Attention Communication and Teaming},
  author={Niu, Yaru and Paleja, Rohan R and Gombolay, Matthew C},
  booktitle={20th International Conference on Autonomous Agents and MultiAgent Systems (AAMAS)},
  pages = {964-973},
  year={2021}
}

@inproceedings{li2021dicg,
  author    = {Sheng Li and Jayesh K. Gupta and Peter Morales and Ross Allen and Mykel J. Kochenderfer},
  title     = {Deep Implicit Coordination Graphs for Multi-Agent Reinforcement Learning},
  booktitle = {20th International Conference on Autonomous Agents and Multiagent Systems (AAMAS)},
  pages     = {764-772},
  year      = {2021}
}

@article{mnih2015dqn,
  title   = {Human-level control through deep reinforcement learning},
  author  = {Mnih, Volodymyr and others},
  journal = {Nature},
  volume  = {518},
  number  = {7540},
  pages   = {529--533},
  year    = {2015}
}

@article{rashid2020monotonic,
  title={Monotonic value function factorisation for deep multi-agent reinforcement learning},
  author={Rashid, Tabish and Samvelyan, Mikayel and De Witt, Christian Schroeder and Farquhar, Gregory and Foerster, Jakob and Whiteson, Shimon},
  journal={Journal of Machine Learning Research},
  volume={21},
  number={178},
  pages={1--51},
  year={2020}
}

@inproceedings{terry2021pettingzoo,
  title={Pettingzoo: Gym for multi-agent reinforcement learning},
  author={Terry, Jordan and Black, Benjamin and Grammel, Nathaniel and Jayakumar, Mario and Hari, Ananth and Sullivan, Ryan and Santos, Luis S and Dieffendahl, Clemens and Horsch, Caroline and Perez-Vicente, Rodrigo and others},
  booktitle={35th International Conference on Neural Information Processing Systems (NIPS)},
  pages={15032--15043},
  year={2021}
}

@article{sutton1988learning,
  title={Learning to predict by the methods of temporal differences},
  author={Sutton, Richard S},
  journal={Machine learning},
  volume={3},
  number={1},
  pages={9--44},
  year={1988},
  publisher={Springer}
}

@inproceedings{ying2018diffpool,
author = {Ying, Rex and You, Jiaxuan and Morris, Christopher and Ren, Xiang and Hamilton, William L. and Leskovec, Jure},
title = {Hierarchical Graph Representation Learning with Differentiable Pooling},
year = {2018},
booktitle = {32nd International Conference on Neural Information Processing Systems (NIPS)},
pages = {4805–4815}
}

@inproceedings{hu2024learning,
  title={Learning Multi-Agent Communication from Graph Modeling Perspective},
  author={Hu, Shengchao and Shen, Li and Zhang, Ya and Tao, Dacheng},
  booktitle={12th International Conference on Learning Representations (ICLR)},
  year={2024}
}

@inproceedings{gerstgrasser2023selectively,
author = {Gerstgrasser, Matthias and Danino, Tom and Keren, Sarah},
title = {Selectively Sharing Experiences Improves Multi-Agent Reinforcement Learning},
year = {2023},
booktitle = {37th International Conference on Neural Information Processing Systems (NIPS)}
}
\bibliographystyle{rlj}

\beginSupplementaryMaterials
\appendix
\section{Environment Details}
\label{sec:supp_envs}

\subsection{MAgent Battle (battle\_v3)}
\label{sec:supp_battle}

\paragraph{Map size to population scale.}
In \textsc{MAgent Battle}, the population scale is determined by the (square) map size. For a map of size $S\times S$, the environment spawns $N$ agents \emph{per team} as listed in Table~\ref{tab:battle_mapsize2n} (e.g., $S{=}40 \Rightarrow 64$ vs $64$).

\begin{table}[h]
\centering
\small
\setlength{\tabcolsep}{6pt}
\renewcommand{\arraystretch}{1.1}
\begin{tabular}{ccccc}
\toprule
$S$ (map size) & 25 & 40 & 45 & 50\\
\midrule
$N$ (agents/team) & 20 & 64 & 81 & 100\\
\bottomrule
\end{tabular}
\caption{Battle scaling: map size $S$ to population $N$ (agents per team).}
\label{tab:battle_mapsize2n}
\end{table}

\paragraph{Agent dynamics and interaction radii.}
Each agent occupies one grid cell ($1{\times}1$), has health $\texttt{hp}=10$, health recovery rate $\texttt{step\_recover}=0.1$, speed $\texttt{speed}=2$, view radius $\texttt{view\_range}=\texttt{CircleRange}(3)$, attack radius $\texttt{attack\_range}=\texttt{CircleRange}(1.5)$, and attack damage $\texttt{damage}=2$.

\paragraph{Rewards.}
The environment provides the following default shaping terms: step reward $\texttt{step\_reward}=-0.005$, death penalty $\texttt{dead\_penalty}=-0.1$, attack penalty $\texttt{attack\_penalty}=-0.1$, and an additional reward of $\texttt{attack\_opponent\_reward}=0.2$ for each successful attack event. In addition, agents receive a kill reward $\texttt{kill\_reward}=5$ upon eliminating an opponent. Episodes run until all \textcolor{blue}{blue} agents are killed or for at most $\texttt{max\_cycles}=200$ steps.

\paragraph{Controlled team and opponent policy.}
In all Battle experiments, we control the \textcolor{red}{red} team, while the \textcolor{blue}{blue} team acts as an opponent using a pretrained IDQN policy, following the same evaluation protocol used by \cite{Li2025ExpoComm}.

\paragraph{Spawn layout.}
At reset, each team is initialized as a dense square formation on opposite sides of the map.
Let $S$ be the map size. The environment sets $\texttt{init\_num}=0.04\,S^2$ and forms a square of side length
$\texttt{side}=2\lfloor \sqrt{\texttt{init\_num}} \rfloor$, placing agents on an every-other-cell lattice (stride 2) to avoid overlap.
The two formations are separated by a horizontal gap of $\texttt{gap}=3$ cells centered around the midline.

\begin{figure}[t]
\centering
\includegraphics[width=0.45\linewidth]{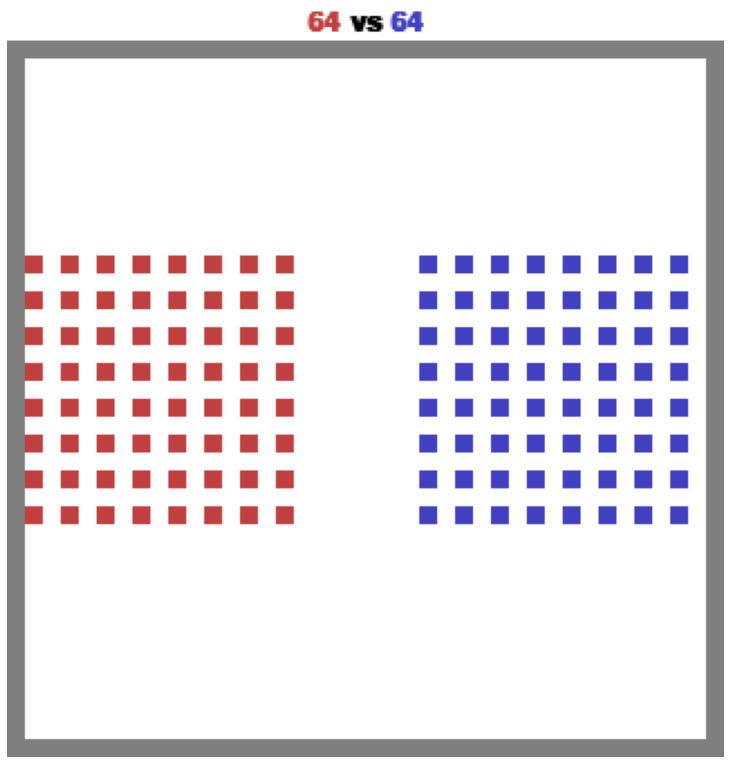}
\caption{Battle reference map at $64$ vs $64$ (map size $40\times 40$). \textcolor{red}{Red} squares denote controlled agents and \textcolor{blue}{blue} squares denote opponent agents.}
\label{fig:supp_battle_ref}
\end{figure}

\paragraph{Reference visualization.}
Figure~\ref{fig:supp_battle_ref} shows a representative $64$ vs $64$ configuration.

\subsection{Pursuit (pursuit\_v3)}
\label{sec:supp_pursuit}

\paragraph{Scenario sizes.}
In \textsc{Pursuit}, the world is a square grid of size $S\times S$, where $S$ depends on the population scale (Table~\ref{tab:pursuit_sizes}).

\begin{table}[h]
\centering
\small
\setlength{\tabcolsep}{8pt}
\renewcommand{\arraystretch}{1.1}
\begin{tabular}{lccccc}
\toprule
Scenario & 20P--8E & 40P--16E & 60P--24E & 80P--32E & 100P--40E \\
\midrule
Map size $S$ & 40 & 45 & 50 & 55 & 60 \\
\bottomrule
\end{tabular}
\caption{Pursuit scaling: pursuers/evaders (P/E) to map size $S\times S$.}
\label{tab:pursuit_sizes}
\end{table}

\paragraph{State and observations.}
The environment maintains a 3-channel grid state consisting of: (i) static obstacles/walls, (ii) pursuer occupancy, and (iii) evader occupancy.
Each pursuer observes a local $R\times R$ patch centered on itself with $\texttt{obs\_range}=R$ (default $R=7$), returned as a 3-channel tensor.
Pursuers act in a discrete action space corresponding to primitive grid movements and a tag action.

\paragraph{Catching rule (surround).}
We use $\texttt{surround}=\texttt{True}$ throughout.
An evader at location $(x,y)$ is removed when pursuers occupy \emph{all open} 4-neighborhood cells (up/down/left/right) around $(x,y)$, and tag the evader.
The number of required surrounding cells accounts for map boundaries and obstacles (i.e., fewer than 4 are required if some neighboring cells are blocked or out-of-bounds).
Each pursuer that participates in a surround event is credited as a catcher.

\paragraph{Rewards and termination.}
The default reward parameters are $\texttt{catch\_reward}=5.0$ (awarded to catchers when an evader is removed), $\texttt{tag\_reward}=0.0$, and $\texttt{urgency\_reward}=0.0$. So, this is a very sparse reward environment.
An episode terminates when all evaders are removed or upon reaching the time limit $\texttt{max\_cycles} = 500$ steps.

\paragraph{Controlled team and opponent policy.}
In all \textsc{Pursuit} experiments, we control the pursuers, while the evaders move randomly around the map.

\paragraph{Initialization.}
At reset, pursuers and evaders are randomly initialized within a (possibly constrained) window of the map. In our runs we use $\texttt{constraint\_window}=1.0$, which corresponds to sampling across the full map.

\begin{figure}[t]
\centering
\includegraphics[width=0.45\linewidth]{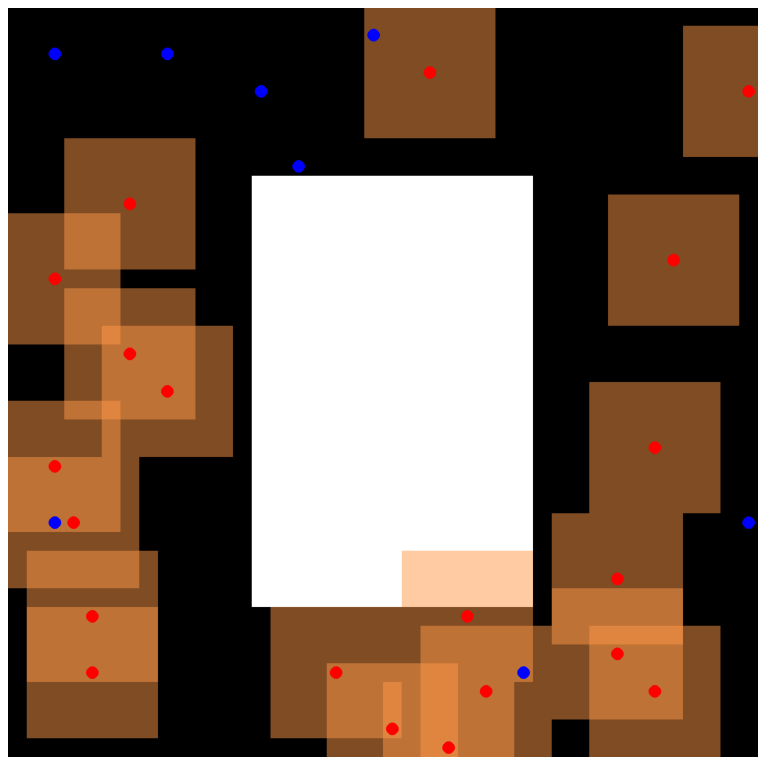}
\caption{Pursuit reference map at 20P--8E (map size $40\times 40$). \textcolor{red}{Red} circles denote pursuers and \textcolor{blue}{blue} circles denote evaders. \textcolor{orange}{Orange} translucent squares indicate local observation windows for pursuers.}
\label{fig:supp_pursuit_ref}
\end{figure}

\paragraph{Reference visualization.}
Figure~\ref{fig:supp_pursuit_ref} shows a representative $20$P--$8$E configuration.

\section{Hyperparameters and Architecture}
\label{sec:hyperparameters}

This section provides a reproducibility-oriented summary of (i) model architectures, (ii) PPO optimization settings, and (iii) communication- and grouping-specific hyperparameters. Unless stated otherwise, the same settings are used across both benchmarks.

\subsection{Model architecture}
\label{sec:hyp_arch}

\paragraph{Actor.}
Each agent uses a recurrent policy with GRU hidden size $h=64$.
At each environment step, the agent observation is first embedded with a linear layer and ReLU, then processed by a GRU:
\[
\text{obs}_t \xrightarrow{\text{Linear+ReLU}} e_t \in \mathbb{R}^{h}
\;\;\xrightarrow{\text{GRU}(h,h)}\;\; u_t \in \mathbb{R}^{h}.
\]
From $u_t$, the actor outputs three heads:
(i) environment action logits, (ii) a binary ``send'' decision logits, and (iii) recipient-selection logits over the $N$ candidate recipients.

\paragraph{Descriptor network.}
A descriptor network encodes per-agent inputs into:
(i) a \emph{grouping descriptor} $z_{\text{grp}}\in\mathbb{R}^{64}$ used for soft group assignments, and
(ii) a \emph{message feature} $z_{\text{msg}}\in\mathbb{R}^{96}$ used as the content representation for communication and for the communication critic.
Descriptor MLPs apply LayerNorm at the output, and observations use running normalization.

\paragraph{Mailbox aggregation (parameter-free attention).}
Incoming messages are aggregated with a single-head dot-product attention in message-feature space:
for each recipient, attention weights are a softmax over senders (scaled by $1/\sqrt{d}$), and the mailbox is the weighted sum of sender message vectors.
All agents are eligible recipients.

\paragraph{Group-aware critic.}
A group critic maps the global state to per-group values using a 2-layer MLP (ReLU, no layer normalization),
\[
V_{\text{grp}}: \;\; s \rightarrow \mathbb{R}^{h}\rightarrow \mathbb{R}^{h}\rightarrow \mathbb{R}^{M},
\]
and produces per-agent values via the (soft) group assignment matrix $P_{\tau}$.

\paragraph{Communication critic.}
We use two small MLPs (ReLU, no layer normalization):
(i) a value network $V(\cdot)$ that takes global state concatenated with an agent feature vector, and
(ii) an edge-value network $Q(i\!\to\!j)$ that takes global state concatenated with sender and recipient features.
The agent feature includes $z_{\text{msg}}$, the incoming mailbox embedding, the send bit, and the $M$-dimensional soft group assignment.

\subsection{Optimization and training}
\label{sec:hyp_ppo}

\paragraph{PPO.}
We use PPO with rollout length 2048 environment steps per update, 8 update epochs, minibatch size 4096, and clipping parameter $\varepsilon=0.2$.
Discounting uses $\gamma=0.99$ and GAE uses $\lambda=0.95$.
The value loss coefficient is 0.5 and the maximum gradient norm is 1.0.
We use Adam with learning rate $3\times 10^{-4}$.
Entropy regularization coefficients are: environment action 0.02, send decision 0.01, and recipient selection 0.01.

\paragraph{Grouping objective and schedules.}
Soft group assignments use a Gumbel--Softmax temperature schedule linearly annealed from $\tau=10$ to $\tau=0.5$ over training.
Assignments are prototype-based using cosine similarity to $M$ learnable prototypes with logit scale 3.0.
Auxiliary grouping losses use: balance weight 0.1, row-entropy weight 0.01, and an edge-utility weight linearly scheduled from 0 to 1 over training.

\paragraph{Macro-step.}
Group assignments and the corresponding group-induced recipient prior are recomputed every $K=10$ environment steps.

\noindent 
For convenience, Table~\ref{tab:shared_hyp_compact} summarizes the shared architecture and training hyperparameters described above.

\begin{table}[h]
\centering
\small
\setlength{\tabcolsep}{7pt}
\renewcommand{\arraystretch}{1.08}
\begin{tabular}{ll}
\toprule
\textbf{Category} & \textbf{Setting} \\
\midrule
\multicolumn{2}{l}{\textit{Network}} \\
Actor hidden size $h$ & 64 (GRU) \\
Descriptor dims & $z_{\text{grp}}:64$, $z_{\text{msg}}:96$ \\
Group critic & MLP: $s\to 64\to 64\to M$ (ReLU) \\
Comm critic & MLPs: $(s||\cdot)\to 64\to 64\to 1$ (ReLU) \\
Normalization & Running obs norm; LayerNorm at descriptor outputs \\
Mailbox aggregation & Single-head dot-product attention, scale $1/\sqrt{d}$ \\
\midrule
\multicolumn{2}{l}{\textit{PPO}} \\
Rollout length & 2048 \\
Minibatch size & 4096 \\
Epochs per update & 8 \\
Clip $\varepsilon$ & 0.2 \\
$\gamma$, GAE $\lambda$ & 0.99, 0.95 \\
Value loss coeff. & 0.5 \\
Entropy coeff. (action / send / recv) & 0.02 / 0.01 / 0.01 \\
Optimizer & Adam, lr $3\times 10^{-4}$ \\
Max grad norm & 1.0 \\
\midrule
\multicolumn{2}{l}{\textit{Grouping / comm}} \\
Macro-step $K$ & 10 \\
Gumbel--Softmax $\tau$ & Linear 10 $\to$ 0.5 \\
Logit scale & 3.0 \\
Aux losses (balance / row-entropy / edge-utility) & 0.1 / 0.01 / Linear 0 $\to$ 1 \\
Mailbox latency & One step ($t\to t{+}1$) \\
\bottomrule
\end{tabular}
\caption{Shared hyperparameters used across Battle and Pursuit.}
\label{tab:shared_hyp_compact}
\end{table}

\subsection{Benchmark-specific settings}
\label{sec:hyp_bench_specific}


\paragraph{Number of groups $M$.}
We set the number of groups using simple scale-dependent heuristics:
\[
\textbf{Battle:}\;\; M=\left\lfloor \sqrt{N}\right\rfloor,
\qquad
\textbf{Pursuit:}\;\; M=\left\lfloor \frac{E}{2}\right\rfloor,
\]
where $N$ is the number of agents \emph{per team} in Battle and $E$ is the number of evaders in Pursuit.
Tables~\ref{tab:groups_battle_reported} and~\ref{tab:groups_pursuit_reported} list $M$ for the population scales reported in the main paper. The choice of these scale-dependent heuristics is further justified through a sensitivity study in Appendix~\ref{sec:sensitivity}.

\begin{table}[h]
\centering
\small
\setlength{\tabcolsep}{10pt}
\renewcommand{\arraystretch}{1.1}
\begin{tabular}{lc}
\toprule
\textbf{Scenario} & \textbf{\# groups $M$} \\
\midrule
\texttt{20v20}   & 4  \\
\texttt{64v64}   & 8  \\
\texttt{81v81}   & 9  \\
\texttt{100v100} & 10 \\
\bottomrule
\end{tabular}
\caption{Battle: number of groups for each scale.}
\label{tab:groups_battle_reported}
\end{table}

\begin{table}[h]
\centering
\small
\setlength{\tabcolsep}{10pt}
\renewcommand{\arraystretch}{1.1}
\begin{tabular}{lc}
\toprule
\textbf{Scenario} & \textbf{\# groups $M$} \\
\midrule
\texttt{20P-8E}    & 4  \\
\texttt{40P-16E}   & 8  \\
\texttt{60P-24E}   & 12 \\
\texttt{80P-32E}   & 16 \\
\texttt{100P-40E}  & 20 \\
\bottomrule
\end{tabular}
\caption{Pursuit: number of groups for each scale.}
\label{tab:groups_pursuit_reported}
\end{table}


\section{Additional Results}
\label{sec:additional-results}

We visualize rollout snapshots from a single illustrative evaluation episode of \textsc{SCoUT} at the headline scale for each benchmark: \textbf{100v100 Battle} (map size $50$, pretrained opponents) and \textbf{100P--40E Pursuit}.
For each episode, we report four frames at \emph{progression} timesteps $t \in \{0, \lfloor T/4 \rfloor, \lfloor T/2 \rfloor, T\}$, where $T$ is the final timestep of the shown episode.
We also report a post-hoc \emph{in-group message fraction} from the same setting to support the claim that the learned interaction structure concentrates communication within groups.

\subsection{Battle (100v100)}
\label{sec:qual_battle}

\begin{figure}[htbp]
  \centering
  \includegraphics[width=0.24\linewidth]{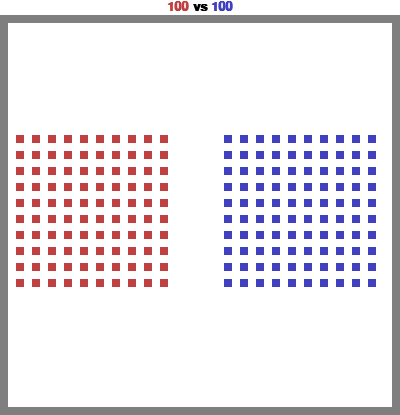}
  \includegraphics[width=0.24\linewidth]{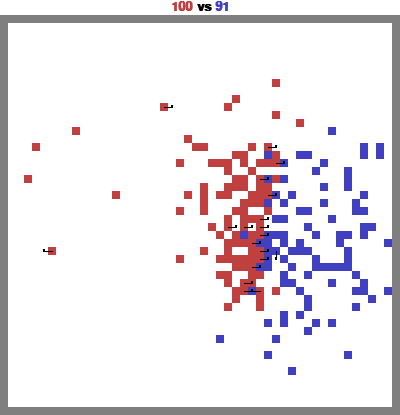}
  \includegraphics[width=0.24\linewidth]{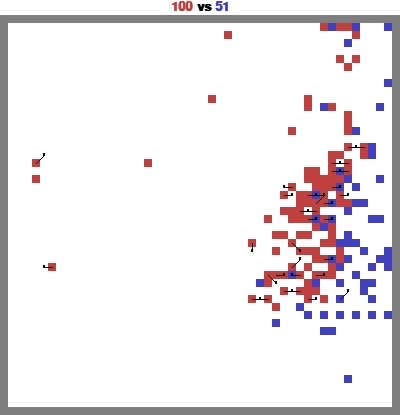}
  \includegraphics[width=0.24\linewidth]{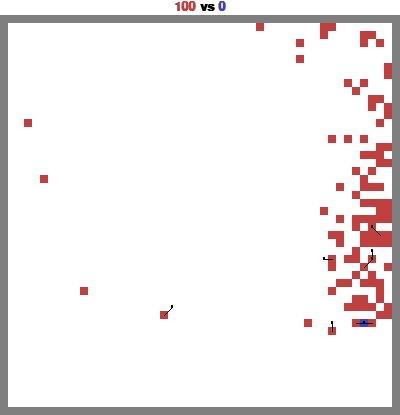}
  \caption{\textsc{Battle} \texttt{100v100}: \textsc{SCoUT} rollout at progression $t = 0, \lfloor T/4 \rfloor, \lfloor T/2 \rfloor, T$ (this episode: $0, 15, 30, 60$ with $T=60$). \textcolor{red}{Red} agents are controlled by \textsc{SCoUT}; \textcolor{blue}{blue} agents are the pretrained IDQN opponent.}
  \label{fig:rollout-battle}
\end{figure}

\paragraph{Key events (Figure~\ref{fig:rollout-battle}).}
At $t=0$, both teams begin in symmetric spawn regions. As contact occurs, \textsc{SCoUT}'s recipient selection becomes structured: messages are preferentially routed within the same (soft) group, which typically corresponds to nearby agents and locally coordinated attacks.
By mid-episode, the controlled \textcolor{red}{red} agents form multiple coherent sub-engagements, rather than a single diffuse front.
By the final frame, remaining \textcolor{red}{red} agents tend to appear in spatially coherent pockets, consistent with localized, group-driven coordination.

\subsection{Pursuit (100P--40E)}
\label{sec:qual_pursuit}

\begin{figure}[htbp]
  \centering
  \includegraphics[width=0.24\linewidth]{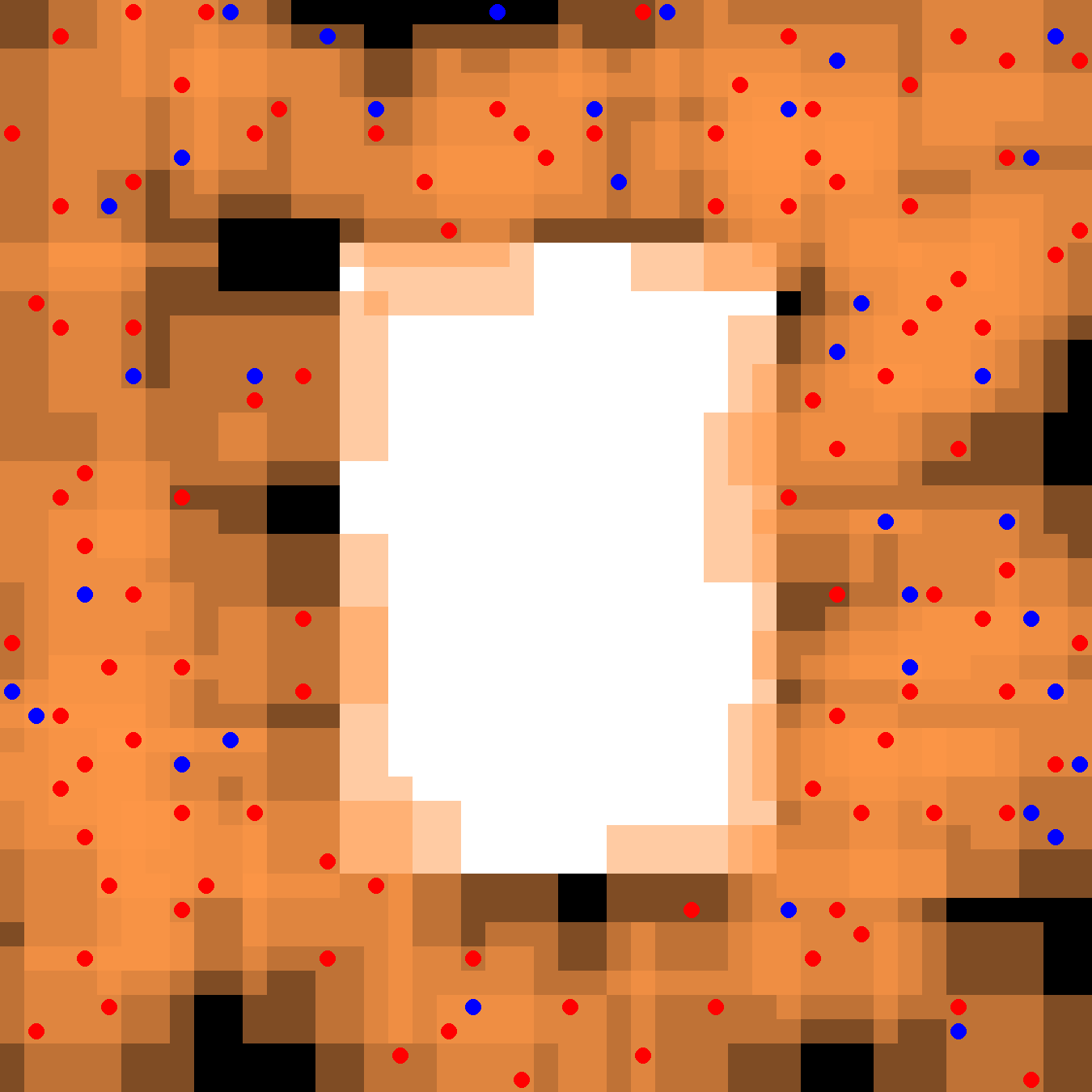}
  \includegraphics[width=0.24\linewidth]{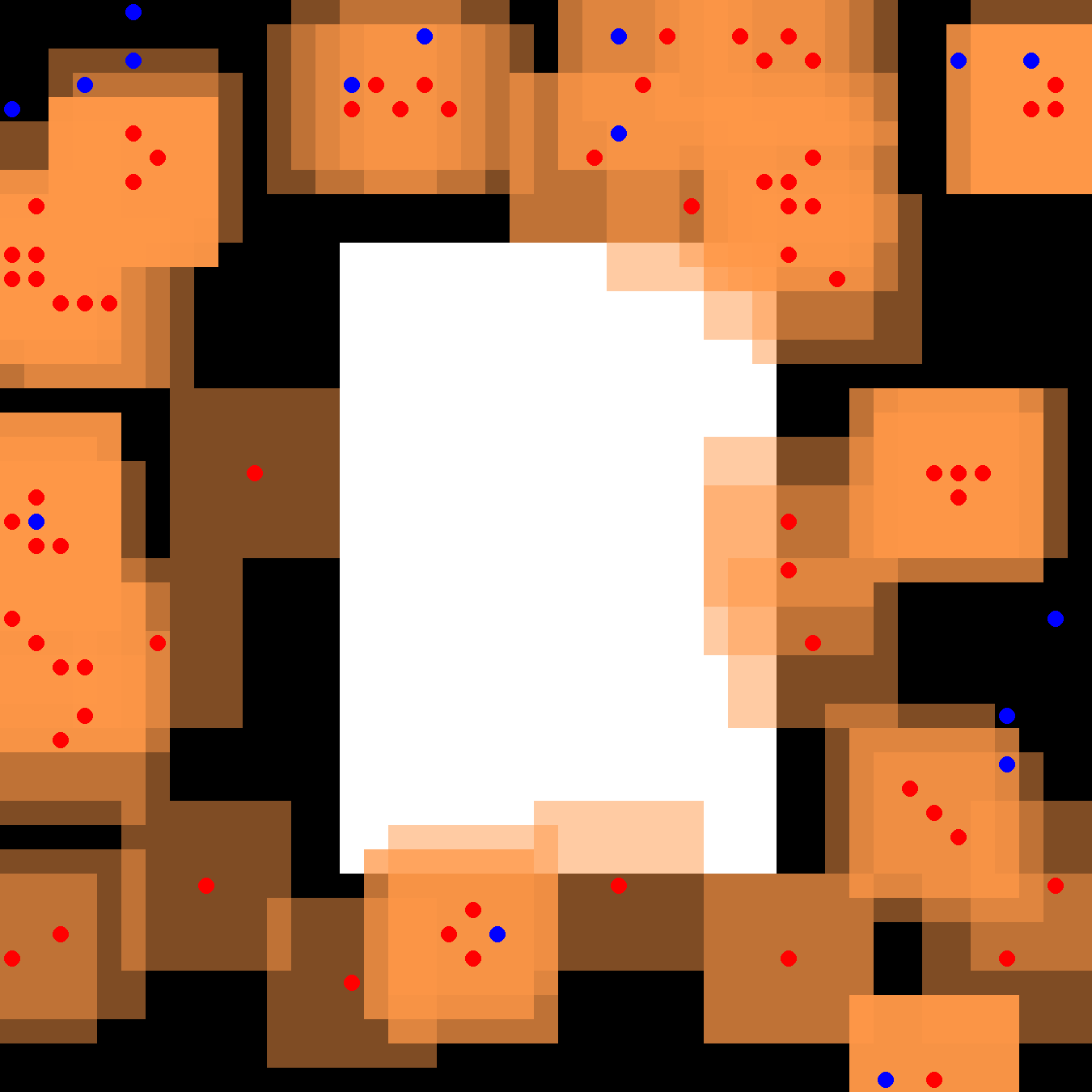}
  \includegraphics[width=0.24\linewidth]{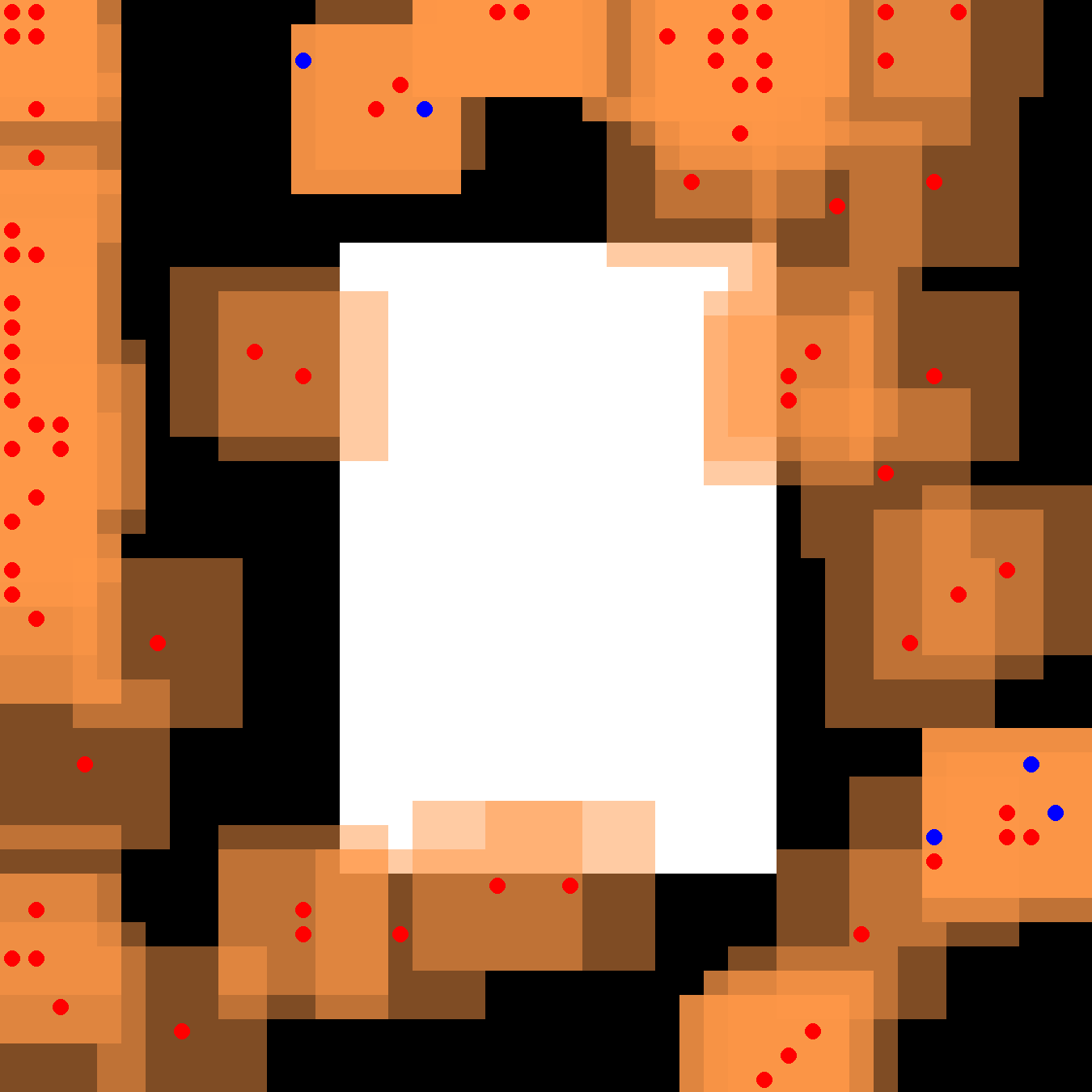}
  \includegraphics[width=0.24\linewidth]{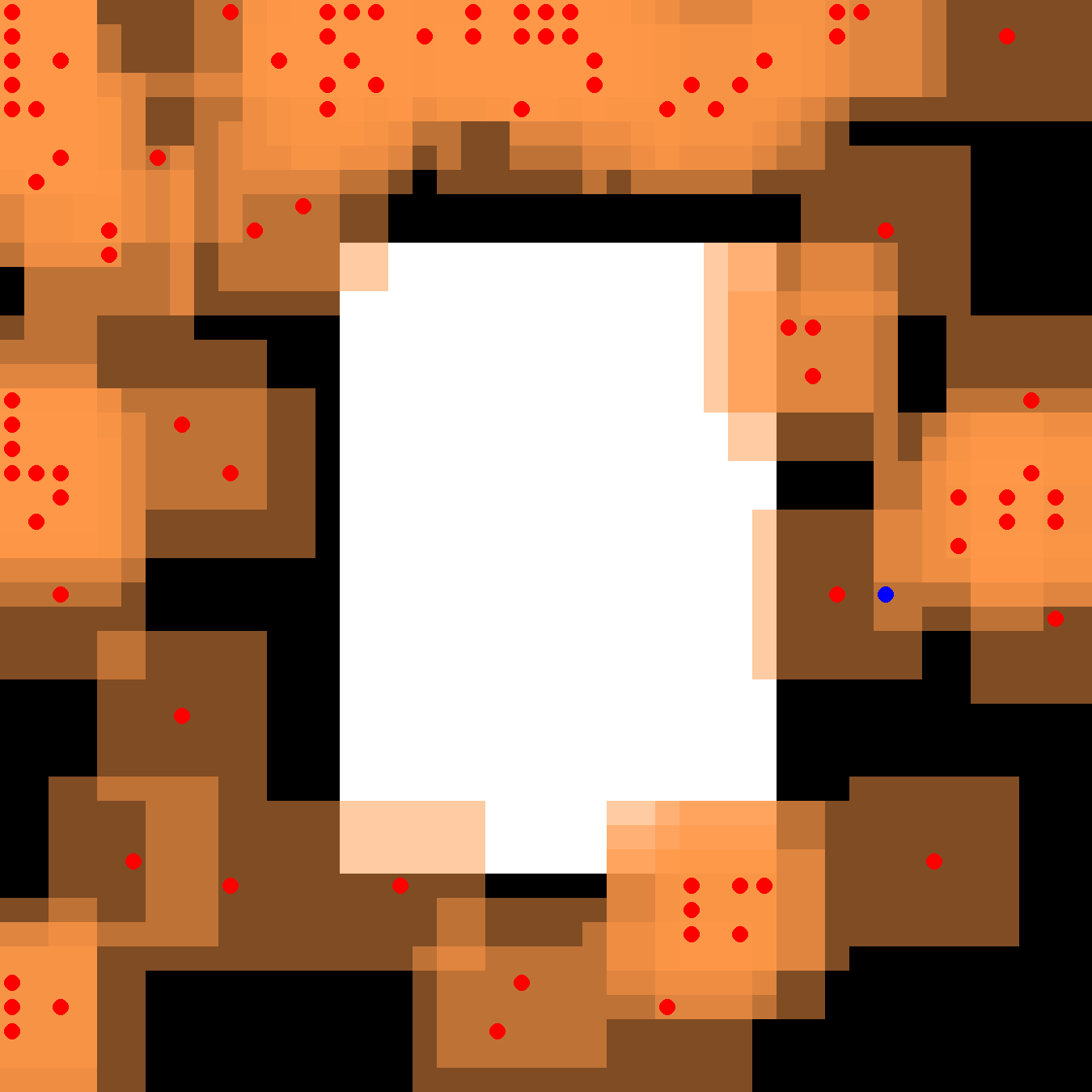}
  \caption{\textsc{Pursuit} \texttt{100P--40E}: \textsc{SCoUT} rollout at progression $t = 0, \lfloor T/4 \rfloor, \lfloor T/2 \rfloor, T$ (this episode: $0, 35, 70, 105$ with $T=135$). \textcolor{red}{Red} circles are pursuers; \textcolor{blue}{blue} circles are evaders. White cells are obstacles/walls; black cells are free space.}
  \label{fig:rollout-pursuit}
\end{figure}

\paragraph{Key events (Figure~\ref{fig:rollout-pursuit}).}
Initially, pursuers spread across the map and begin forming local coalitions around nearby evaders.
As the episode progresses, subteams emerge that repeatedly execute surround-and-capture behaviors in different regions of the map (visible as clusters of \textcolor{red}{pursuers} tightening around \textcolor{blue}{evaders}).
This behavior is consistent with a group-induced recipient bias: communication emphasizes coordination among agents that are already acting as a local capture unit, reducing unnecessary cross-map broadcast.

\subsection{Effect of grouping: in-group message fraction}
\label{sec:ingroup-frac}

To quantify whether communication aligns with the learned interaction structure, we compute the \emph{in-group message fraction}: the fraction of realized sender--recipient message edges $(i\!\to\!j)$ for which sender and recipient belong to the same group.
Concretely, at each macro-step we obtain soft assignments $P_{\tau}\in[0,1]^{N\times M}$ and define a hard group label $\hat{g}(i)=\arg\max_{m} P_{\tau}(i,m)$.
Let $\mathcal{E}$ denote the set of all realized sender-recipient message edges during an episode. We report
\[
\text{InGroupFrac} \;=\; \frac{1}{|\mathcal{E}|}\sum_{(i\to j)\in \mathcal{E}} \mathbb{I}\!\left[\hat{g}(i)=\hat{g}(j)\right].
\]
We average this metric over 10 evaluation seeds. This is a diagnostic metric used only for analysis and is not itself optimized as a training objective.
Table~\ref{tab:ingroup_all_scales} reports the in-group fraction across all scales, together with the random baseline $1/M$.

\begin{table}[t]
\centering
\scriptsize
\setlength{\tabcolsep}{2.6pt}
\renewcommand{\arraystretch}{1.15}
\newcommand{\unc}[1]{{\tiny$\pm$#1}}
\resizebox{\linewidth}{!}{%
\begin{tabular}{lcccc|ccccc}
\toprule
\rowcolor{tablegray}
\textbf{Metric} &
\textbf{20v20} & \textbf{64v64} & \textbf{81v81} & \textbf{100v100} &
\textbf{20P-8E} & \textbf{40P-16E} & \textbf{60P-24E} & \textbf{80P-32E} & \textbf{100P-40E} \\
\midrule

\multicolumn{5}{l}{\textsc{Battle}} &
\multicolumn{5}{l}{\textsc{Pursuit}} \\
\textbf{In-group frac.} &
0.343\unc{0.046} & 0.257\unc{0.009} & 0.488\unc{0.032} & 0.406\unc{0.010} &
0.297\unc{0.024} & 0.319\unc{0.010} & 0.376\unc{0.056} & 0.238\unc{0.029} & 0.397\unc{0.003} \\
\textbf{Random baseline ($1/M$)} &
0.250 & 0.125 & 0.111 & 0.100 &
0.250 & 0.125 & 0.083 & 0.062 & 0.050 \\
\textbf{\# groups ($M$)} &
4 & 8 & 9 & 10 &
4 & 8 & 12 & 16 & 20 \\

\bottomrule
\end{tabular}%
}
\caption{\textbf{In-group message fraction across scales (10 evaluation seeds).}
We report mean $\pm$ std of the in-group message fraction (Eq.~above) computed over realized sender--recipient message edges.
The random baseline corresponds to uniformly random group membership, which yields expected in-group fraction $1/M$.}
\label{tab:ingroup_all_scales}
\vspace{-3mm}
\end{table}

Across all reported scales, the in-group message fraction is well above the $1/M$ baseline, indicating that recipient selection is strongly aligned with the learned group structure.
This supports the qualitative rollouts in Figure~\ref{fig:rollout-battle} and Figure~\ref{fig:rollout-pursuit} and is consistent with the role of grouping in \textsc{SCoUT}: it induces a slowly-varying interaction structure during training that reduces diffuse broadcast and promotes localized coordination.


\section{Sensitivity to Number of Groups and Macro-Step Length}
\label{sec:sensitivity}

\textsc{SCoUT} introduces two environment-facing hyperparameters that control the \emph{granularity} and \emph{temporal persistence} of the learned interaction structure: the number of groups $M$ and the macro-step length $K$. This section evaluates how sensitive performance is to $(M,K)$, and whether the choice of $K{=}10$ and scale-dependent $M$, used in the main experiments, lies in a stable high-performing region.

\paragraph{Settings.}
We perform controlled sweeps on two representative, computationally manageable scales:
\textsc{Pursuit} \texttt{40P--16E} and \textsc{Battle} \texttt{64v64}.
These settings are large enough to exhibit nontrivial coordination dynamics, while being substantially cheaper computationally than the largest stress-test cases. Even at these scales, each $4{\times}4$ sweep required approximately \textbf{4 days} per benchmark, making analogous sweeps at larger scales prohibitively expensive.

\subsection{Experimental setup}
\label{sec:sensitivity_setup}

\paragraph{Grid over $(M,K)$.}
For each benchmark, we evaluate a $4\times4$ grid:
\[
M \in \{4,6,8,12\},
\qquad
K \in \{5,10,15,20\}.
\]
The grid includes the default configuration used in the main paper, $(M{=}8, K{=}10)$, which corresponds to our scale rules at \texttt{64v64} Battle ($M=\lfloor\sqrt{64}\rfloor=8$) and \texttt{40P--16E} Pursuit ($M=\lfloor 16/2\rfloor=8$).

\begin{figure}[htbp]
  \centering
  \includegraphics[width=0.48\linewidth]{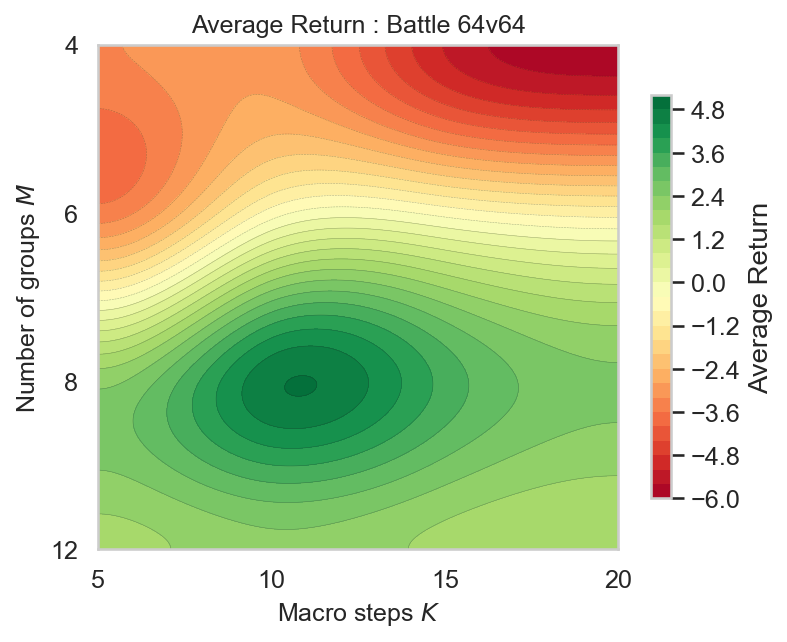}
  \includegraphics[width=0.48\linewidth]{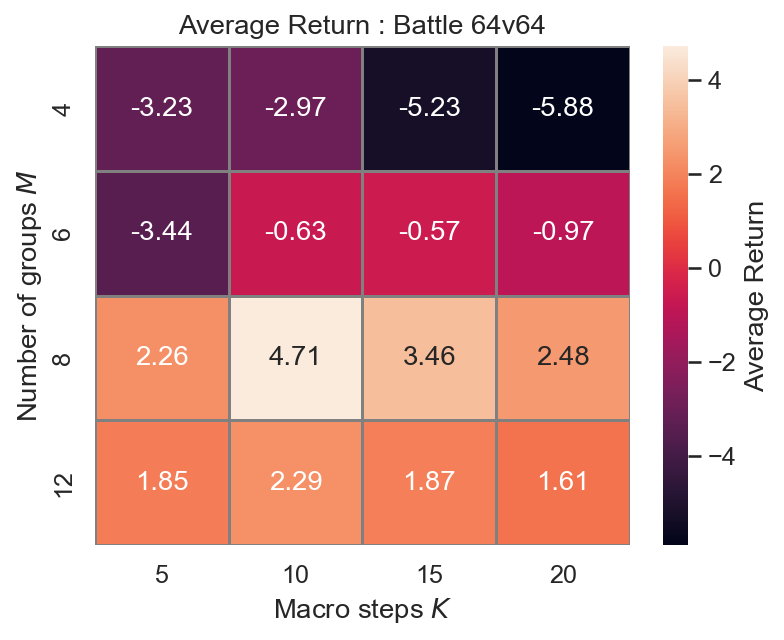}
  \caption{Sensitivity of \textsc{SCoUT} to $(M,K)$ on \textsc{Battle} \texttt{64v64}.
  \textbf{Left:} Interpolated contour plot.
  \textbf{Right:} Heatmap with exact returns for each evaluated $(M,K)$ pair.}
  \label{fig:sensitivity-battle}
\end{figure}

\begin{figure}[htbp]
  \centering
  \includegraphics[width=0.48\linewidth]{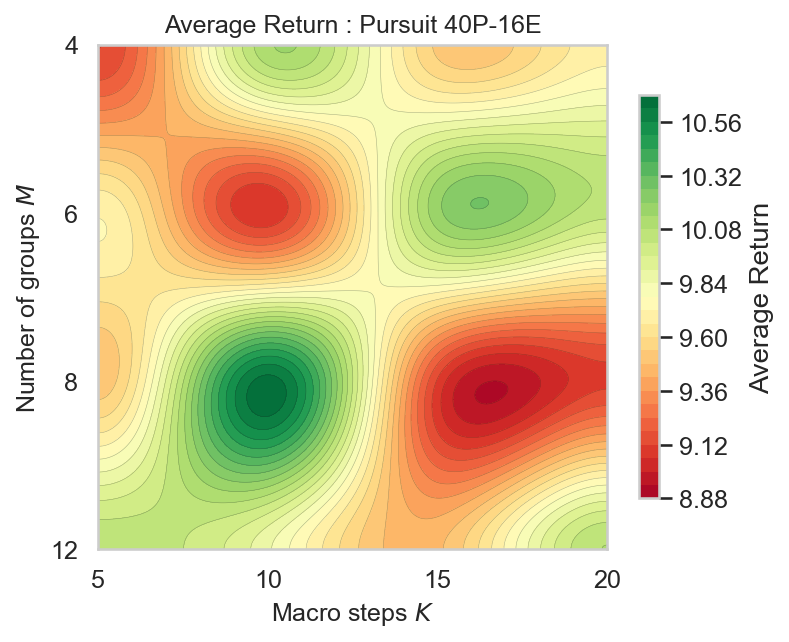}
  \includegraphics[width=0.48\linewidth]{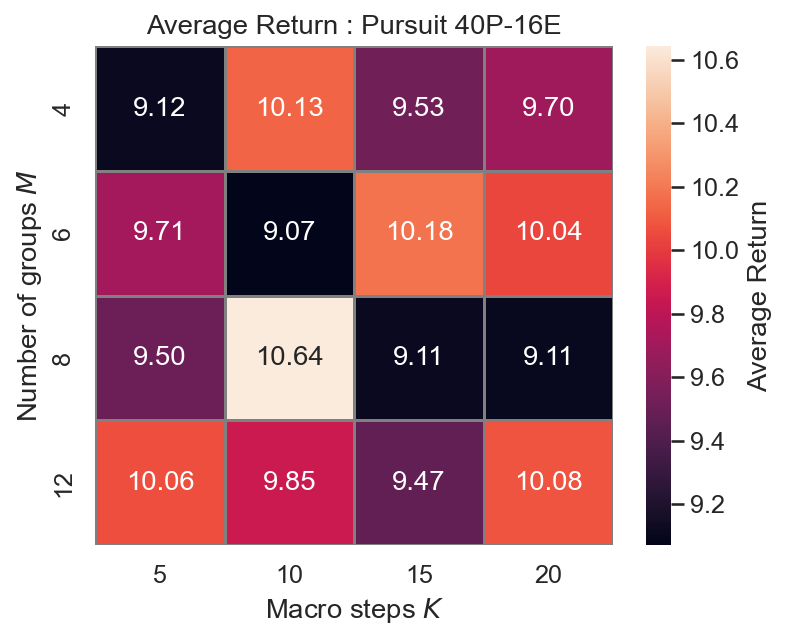}
  \caption{Sensitivity of \textsc{SCoUT} to $(M,K)$ on \textsc{Pursuit} \texttt{40P--16E}.
  \textbf{Left:} Interpolated contour plot for the coarse trend (higher is better).
  \textbf{Right:} Heatmap with exact returns for each evaluated $(M,K)$ pair.}
  \label{fig:sensitivity-pursuit}
\end{figure}

\paragraph{Training protocol and metric.}
All runs use identical hyperparameters except for $(M,K)$ (Appendix~\ref{sec:hyperparameters}) and are trained for 2000 iterations.
We report the final average episode return for each configuration.
Figures~\ref{fig:sensitivity-battle} and~\ref{fig:sensitivity-pursuit} visualize the resulting performance surfaces.

\subsection{Results and interpretation}
\label{sec:sensitivity_interpretation}

\paragraph{Battle \texttt{64v64}.}
Figure~\ref{fig:sensitivity-battle} shows a sharp dependence on $M$.
Very small $M$ performs poorly across $K$, with $M{=}4$ yielding the lowest returns for all $K$.
In contrast, intermediate group counts are substantially stronger: the best configuration is $(M{=}8, K{=}10)$, with nearby settings remaining competitive.
For large $M$, average reward remains positive but is typically below the best region.
This pattern is consistent with the intuition that Battle benefits from meaningful group structure, while still retaining enough within-group mass to coordinate local engagements.

\paragraph{Pursuit \texttt{40P--16E}.}
Figure~\ref{fig:sensitivity-pursuit} shows a clear best-performing configuration at $(M{=}8, K{=}10)$.
Furthemore, several settings also achieve competitive return: $(M{=}6,K{=}15)$ at $10.18$, $(M{=}4,K{=}10)$ at $10.13$, and $(M{=}12,K{=}20)$ at $10.08$).
Overall, the sweep supports the default $(8,10)$ while indicating that \textsc{SCoUT} is less sensitive to the choice of $(M,K)$in \textsc{Pursuit} as compared to \textsc{Battle}, which showed a more structured and sharper dependence on $M$.

\paragraph{Takeaway.}
Across both benchmarks, the sweeps place the best-performing region at or near the default choice $(M{=}8,K{=}10)$ used in the main experiments. The observed trends align with the roles of the two hyperparameters:
$M$ controls interaction granularity---too small can reduce selectivity and induce broadcast-like communication; too large can fragment coalitions. On the other hand, $K$ controls {persistence}---too small can introduce unnecessary churn; too large can make the interaction structure stale in dynamic settings.
Together, these results provide additional justification for using $K{=}10$ with scale-dependent $M$ in the main paper without per-scale tuning.

\end{document}